\documentclass[useAMS,usenatbib]{mn2e}
\usepackage{color}
\usepackage{graphicx}
\usepackage{rotating}
\usepackage{lscape}
\usepackage{amssymb,amsmath}
\usepackage{url}
\usepackage{float}

\title[Cepheid multi-wavelength Fourier parameters]
{On the Variation of Fourier Parameters for Galactic and LMC Cepheids at Optical, Near-Infrared and Mid-Infrared Wavelengths}
\author[Bhardwaj et al.]{Anupam Bhardwaj$^1$\thanks{E-mail:
anupam.bhardwajj@gmail.com}, Shashi M. Kanbur$^2$, Harinder P. Singh$^1$, Lucas M. Macri$^3$,
\newauthor 
Chow-Choong Ngeow$^4$ \\
1. Department of Physics \& Astrophysics, University of Delhi, Delhi 110007, India. \\
2. State University of New York, Oswego, NY 13126, USA.\\
3. Mitchell Institute for Fundamental Physics \& Astronomy, Department of Physics \& Astronomy, Texas A\&M University, \\
~~~College Station, TX 77843, USA \\
4. Graduate Institute of Astronomy, National Central University, Jhongli 32001, Taiwan \\
}
\begin{document}

\date{Accepted 2014 December 15.  Received 2014 December 15; in original form 2014 July 29}

\pagerange{\pageref{firstpage}--\pageref{lastpage}} \pubyear{2014}

\maketitle

\label{firstpage}

\begin{abstract}
  We present a light curve analysis of fundamental-mode Galactic and Large Magellanic Cloud (LMC)
  Cepheids based on the Fourier decomposition technique.  We have compiled light
  curve data for Galactic and LMC Cepheids in optical ({\it VI}), near-infrared
  ({\it JHK}$_s$) and mid-infrared (3.6 $\&$ 4.5-$\mu$m) bands from the
  literature and determined the variation of their Fourier parameters as a
  function of period and wavelength. We observed a decrease in Fourier amplitude
  parameters and an increase in Fourier phase parameters with increasing
  wavelengths at a given period. We also found a decrease in the skewness and acuteness 
  parameters as a function of wavelength at a fixed period.
  We applied a binning method to analyze the
  progression of the mean Fourier parameters with period and wavelength.
  We found that for periods longer than about 20 days, the values of the
  Fourier amplitude parameters increase sharply for shorter wavelengths 
  as compared to wavelengths longer than the $J$-band.
  We observed the variation of the Hertzsprung progression with wavelength.  
  The central period of the Hertzsprung progression was found to increase with
  wavelength in the case of the Fourier amplitude parameters and decrease
  with increasing wavelength in the case of phase parameters. We also observed a small
  variation of the central period of the progression between the Galaxy and LMC,
  presumably related to metallicity effects. These results will provide useful
  constraints for stellar pulsation codes that incorporate stellar atmosphere models
  to produce Cepheid light curves in various bands.

\end{abstract}

\section{Introduction}

Cepheid variables are bright, population I periodic radial pulsators that
exhibit regular light curves and obey a well known Period-Luminosity
relation \citep{levitt12}, that is an important tool in the extra-galactic distance scale. Fourier analysis
methods have been used extensively to describe Cepheid light curve structure and its 
variation with period. In particular, the amplitude ratios ($R_{21}$ and $R_{31}$) and
phase differences ($\phi_{21}$ and $\phi_{31}$) have been used to quantitatively 
describe the progression of Cepheid light curve shape with period \citep{simona77}. 
The Fourier decomposition method was further revived by \citet{slee81}, who
used a sample of 57 Cepheids and discussed the variation of Fourier parameters 
with period. The sharp breaks in the progressions of Fourier 
parameters with period, occurring near 10 days, were attributed to the 
resonance $P_{2}/P_{0}=0.5$, in the normal mode spectrum \citep{simonsch76, simona77, slee81}. 
Later, the method was used extensively by \citet{simont82} to analyze 
the progressions of Fourier parameters and light curve structure of a large sample of 
field RR Lyrae stars. The light and velocity curves were Fourier decomposed to 
compare with theoretically modeled light curves 
\citep{sdavis83, simon85}. Similar studies of the light curve structures 
of RR Lyrae variables were carried out by 
\citet{simona85, kovacs86}. The Fourier phase parameter ($\phi_{31}$) was used in empirical relations to determine
the metallicity of fundamental mode RR Lyrae stars \citep{JK96}. 
The studies on theoretical light curves of Cepheid variables
using the skewness and acuteness parameters together with the variation of Fourier parameters were carried out by 
\citet{stelling86, stelling87, bonohp}. The central period of Hertzsprung progression was also
determined using the Fourier parameters \citep{mosk92, welchhp97, beauli98} and skewness/acuteness parameters \citep{bonohp}.   
Other studies employing the Fourier
decomposition technique to analyze the light curves of Cepheid variables, are \citet{simona86,
antone86, ander87, ander88, simona88, poretti94, simona95, stetson96, welchhp97, beauli98}.

Recent applications of this method include the reconstruction of 
Cepheid light curves with Fourier technique \citep{ngeow03}, 
and classification of variable star light curves based on Fourier parameters
and Principal Component Analysis \citep[PCA,][]{deb09}. 
Most of the recent studies on Fourier decomposition involve the determination of 
physical parameters like absolute magnitude, metallicity, effective temperature, 
luminosity for RR Lyrae variables \citep{deb10, nemec11}. The Fourier decomposition
technique has been further extended to describe the chemical and structural properties of the LMC \citep{deb14}.

In this work, we analyze the light curves of fundamental-mode Galactic and LMC Cepheids in 
multiple bands using Fourier decomposition techniques.
In Section~\ref{sec:fourier}, we provide a brief description of the 
application of the Fourier decomposition method. 
In Section~\ref{sec:lc_data}, we discuss the Galactic and LMC Cepheid 
light curve data compiled from the literature for optical, 
near-infrared and mid-infrared wavelengths.
In Section~\ref{sec:fparams}, we describe the application of Fourier 
decomposition to Galactic and LMC Cepheid light curves. 
Further, we discuss the variation of Fourier parameters with period in 
each band separately. In 
Section~\ref{sec:comp_fparams}, we compare the Fourier parameters 
in multiple bands and comment on their progression with period and
wavelength. We also discuss the variation of mean Fourier parameters together with skewness and acuteness parameters 
as a function of wavelength at a given period. In Section~\ref{sec:hp_logp}, we summarize the variation
of the central period of the Hertzsprung progression with wavelength for each Fourier parameter in the Galaxy 
and LMC. A discussion on our results and important 
conclusions arising from this study are presented in Section~\ref{sec:discuss}. 

Our results will provide important constraints for stellar pulsation codes that incorporate 
stellar atmosphere models to produce wavelength-dependent theoretical Cepheid light curves.


\section{Fourier Decomposition Technique}
\label{sec:fourier}

Fourier decomposition is a robust method to study the light curves of variable stars. 
This method was revived and refined by \citet{slee81} in its modern form. They described 
how the lower order Fourier coefficients can completely describe the structure of the 
light curve. The Fourier coefficients and Fourier parameters
are now widely used to derive empirical relations to determine physical parameters of 
variable stars, in particular for fundamental mode RR Lyrae stars.

In this study we have used a sine Fourier series to fit the multi-band 
light curves of Galactic and LMC Cepheids, 

\begin{equation}
m(t) = m_{0}+\sum_{i=1}^{N}A_{i} \sin(i \omega (t-t_{0}) + \phi_{i}),
\label{eq:sinfit0}
\end{equation}

\noindent where $m(t)$ is the observed magnitude, $m_{0}$ is the mean magnitude from the Fourier fit,
$t$ is the time of observation, $\omega = 2\pi/P$ is the
angular frequency and $t_{0}$ corresponds to the epoch of maximum brightness. In this study, we have taken $t_{0}$ 
as the time of minimum magnitude from the light curve data for each Cepheid, which is used to obtain a phased
light curve that has maximum light at phase zero. $A_{i}$ and $\phi_{i}$ are amplitude and
phase coefficients respectively. Since the period \emph{P} is known, the light curves are phased using
\begin{equation*}
x = \mathrm{frac}\left(\frac{t-t_{0}}{P}\right).
\end{equation*}

 Since the values of $x$ range from 0 to 1, corresponding to a full cycle of pulsation, equation~(\ref{eq:sinfit0}) can be written as:

\begin{equation}
m = m_{0}+\sum_{i=1}^{N}A_{i} \sin(2 \pi i x + \phi_{i}).
\label{eq:foufit1}
\end{equation}

Here, \emph{N} is the optimum order of fit, which is generally chosen depending on the size of least square residuals.
Furthermore, coefficients A$_{1}$\dots A$_{\emph{N}}$ and $\phi_{1}\dots \phi_{\emph{N}}$ are
extracted from the fit to give Fourier parameters,

\begin{equation}
R_{i1} = \frac{A_{i}}{A_{1}} ; \phi_{i1} = \phi_{i} - i\phi_{1},
\label{eq:fparams} 
\end{equation}

\noindent where $i > 1$. The $\phi_{i1}$ are generally adjusted to lie between 0 and $2\pi$. 
The errors in the derived Fourier parameters are determined using the propagation of errors in the coefficients \citep{deb10}.

\section{The data}
\label{sec:lc_data}

\begin{table*}
\begin{center}
\caption{The Galactic and LMC Cepheid multi-wavelength light curve data selected for the present analysis.} 
\label{tab:data_table}
\scalebox{1.0}{
\begin{tabular}{|p{2.cm}|p{2.cm}|p{3.cm}|p{3.5cm}|p{4cm}|}
\hline
\hline

{\bf Band} 	& 		\multicolumn{2}{c}{\bf Galaxy}~~~~~~~~~~~~~	& \multicolumn{2}{c}{\bf LMC}~~~~~~~~~~   \\
\hline
\hline
  \centering   	& No. of stars & References   				& ~~~~~~~~~~~~~~No. of stars & References	\\
\hline
\hline
$V$		&	447	&\citet{berdi08}				& ~~~~~~~~~~~~~~1832	&\citet{sosz2008, ulac13} \\
$I$		&	351	&-						& ~~~~~~~~~~~~~~1844	&-			\\
\hline
$J$		&	186	&\citet{welch84, laney92, barnes97, monson11}	& ~~~~~~~~~~~~~~474	& \citet{macri14, persson04} \\
$H$		&	186	&  -               				& ~~~~~~~~~~~~~~532	&  -\\
$K_{\rm s}$	&	186	& -                				& ~~~~~~~~~~~~~~488	& -\\
\hline
3.6-$\mu m$	&	37	&\citet{monson12}				& ~~~~~~~~~~~~~~84	&\citet{vicky11} \\
4.5-$\mu m$	&	37	&-						& ~~~~~~~~~~~~~~84	& -\\
\hline
\hline
\end{tabular}}
\end{center}
\end{table*}

The data selected for present analysis is described in Table~\ref{tab:data_table}. A brief description of
each catalogue/source used is presented in the following subsections.

\subsection{Optical Wavelengths}

\subsubsection{Galactic Cepheids}
The light curve data for Galactic Cepheids in the Johnson $V$- and Kron-Cousins $I$-bands 
were extracted from the catalogue of \citet{berdi08}. This catalogue
gathers photoelectric observations of Galactic Cepheids made between 1986 and
2004 by Berdnikov and his collaborators in a series of papers \citep{berdi87,
berdi92, berdi93, berdit95, berdi98, berdit01, berdit03, berdit04}.  Our
analysis makes use of 447 and 351 Galactic Cepheids with data in the $V$- and $I$-bands, 
respectively.  Since this catalogue is compiled from a series of
observations carried out over two decades, the number of data points for each
star ranges from 20 to nearly 400. The periods of the variables were extracted
from the database of Galactic Classical Cepheids \citep{fernie95}.

\subsubsection{LMC Cepheids}
The light curve data for LMC Cepheids in the $V$- and $I$-bands were taken from the
third phase of the Optical Gravitational Lensing Experiment (OGLE-III) survey
\citep{sosz2008}. The observations were carried out using a dedicated 1.3-m
telescope at the Las Campanas Observatory, Chile. Our analysis makes use of 1806
and 1818 light curves in the $V$- and $I$-bands, respectively. The light curves are
fairly well sampled with a large number of data points in $I$-band, and covering nearly full phase in
both optical bands. We also make use of the period and initial epoch provided in
this database to Fourier fit the light curves. 

We also extracted the light curve data in the $V$- and $I$-bands for 26 long period Cepheids from
OGLE-III Shallow Survey in the LMC \citep{ulac13}. The photometric data for these Cepheids were also collected using
the 1.3-m Warsaw Telescope located at Las Campanas Observatory. Since the photometric system is exactly similar in both the
surveys (\citet{sosz2008} $\&$ \citet{ulac13}), we increase our sample to have 1832 Cepheids in $V$-band and 1844 Cepheids in $I$-band. 

\subsection{Near-Infrared Wavelengths}

\subsubsection{Galactic Cepheids}
We compiled photometric data for 186 Galactic Cepheids in the $JHK_{\rm s}$ bands with
full phase coverage using several sources in the literature \citep{welch84,
laney92, barnes97, monson11}.  The light curves taken from \citet{monson11}
for 129 Galactic Cepheids were obtained during a span of 10 months in 2008 using
the BIRCAM instrument at the 0.6-m telescope of the University of Wyoming Red
Buttes Observatory (RBO). These light curves have an average of 22 observations
per star, providing reasonable phase coverage. We included 41 light curves from
\citet{laney92} that were obtained between 1982 to 1990 at the Sutherland
observing station of South African Astronomical Observatory.  Most of the
observations were carried out using the 0.75-m telescope and the Mark II
infrared photometer and the remaining observations were made using the 1.9-m
telescope and the Mark III infrared photometer.  These light curves have 31 data
points per star. We also made use of 8 light curves from \citet{barnes97},
obtained at the 1.3-m telescope at Kitt Peak National Observatory using the OTTO
and SQIID instruments. Lastly, we incorporated 8 variables from \citet{welch84}
that were obtained using the 1-m Swope telescope at Las Campanas Observation,
Chile and the 0.6-m telescope at Mount Wilson using the 0.6-m reflector.

Since these near-infrared light curve data were obtained by the various authors
using different photometric systems, we transformed them into the 2MASS system
using the transformations provided as part of their all-sky data 
release\footnote{\url{http://www.ipac.caltech.edu/2mass/releases/allsky/doc/sec6_4b.html}}.

\begin{figure}
\begin{center}
\includegraphics[width=0.5\textwidth,keepaspectratio]{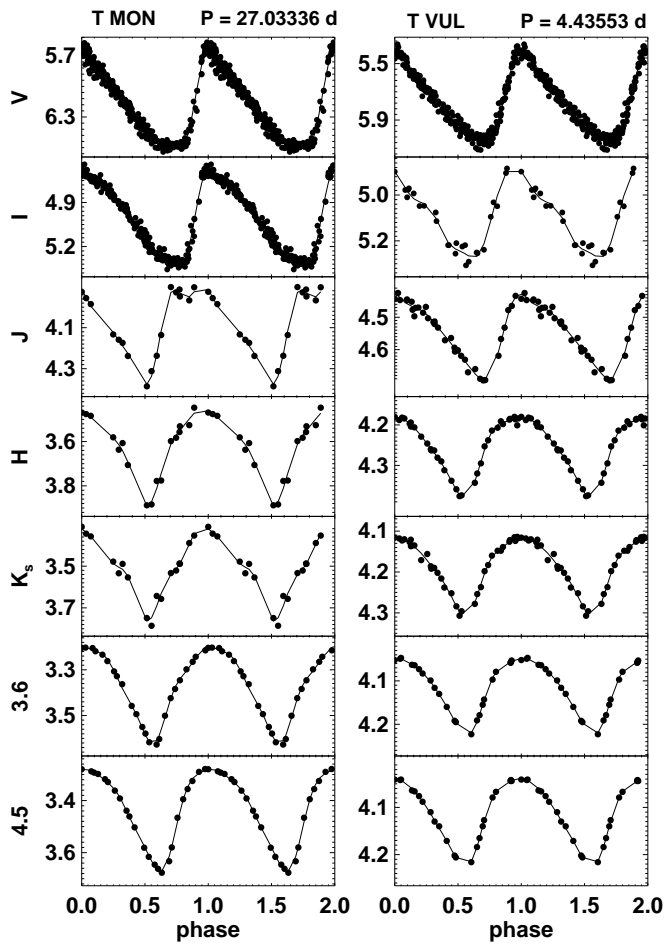}
\end{center}
\caption{Examples of the Fourier-fitted light curves of two Galactic Cepheids in multiple
  bands. The star ID is given at the top of each panel.}
\label{fig:galactic_lc.eps}
\end{figure}

\begin{figure}
\begin{center}
\includegraphics[width=0.5\textwidth,keepaspectratio]{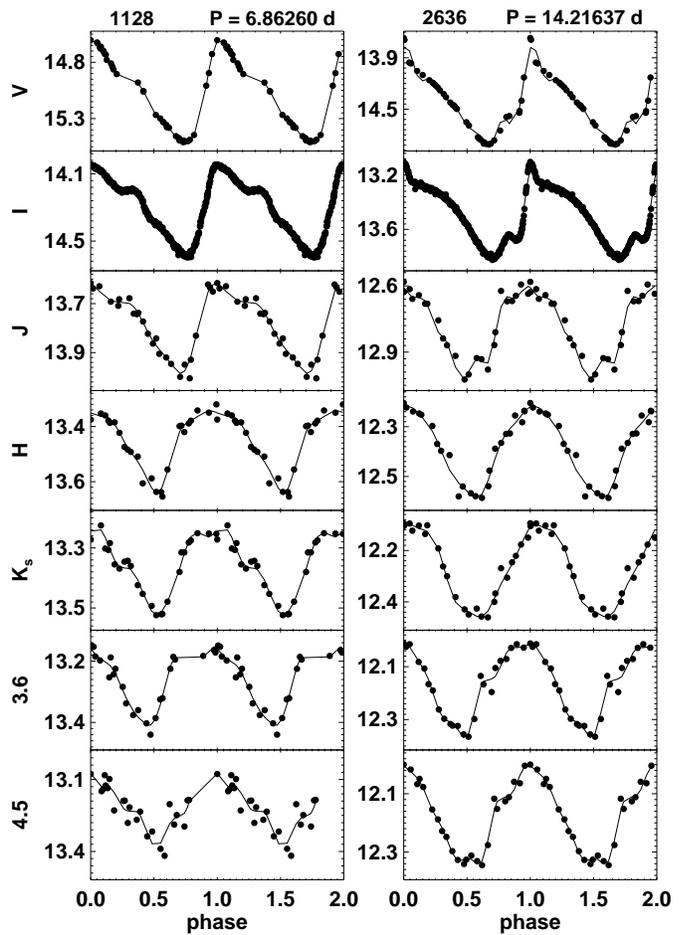}
\end{center}
\caption{Same as Figure 1, but for two LMC Cepheids. Star ID's consisting of 4 numbers are based on the OGLE-III catalog. 
These variables are identified as HV2337 \& HV1006 in \citet{persson04} and \citet{vicky11}}.
\label{fig:lmc_lc.eps}
\end{figure}

\subsubsection{LMC Cepheids}
Our analysis made use of combined LMC near-infrared light curve data for 474,
532 and 488 Cepheids in $J$, $H$ and $K_{\rm s}$, respectively, from the two sources
listed below.

We used the light curve data from \citet{macri14}, who carried out
a $JHK_{\rm s}$ survey of the central $\sim 18 \sq ^{\circ}$~of the LMC using the
CPAPIR camera at the Cerro Tololo Inter-American Observatory (CTIO) 1.5-m
telescope, operated as part of the SMARTS consortium. The variables, originally
identified by OGLE-III \citep{sosz2008} range in period from 1 to 37 days, with
an average number of 16 phase points per object. The observations were
calibrated by the authors into the 2MASS system. We used 384, 442, 398
light curves in $J$, $H$ \& $K_{\rm s}$, respectively.

We have also used the $JHK_{\rm s}$ data for 90 Cepheids in the LMC from
\citet{persson04}. These observations were carried out with the 1-m Swope and
2.5-m duPont telescopes at Las Campanas Observatory between 1993 to 1997. These
stars have periods in the range of 2 to 134 days and an average of 22
observations per band.  These observations were reported using the LCO
photometric system, and were transformed into the 2MASS system using the
previously referenced relations.

\subsection{Mid-Infrared Wavelengths}

We used the 3.6 and 4.5-$\mu$m light curves of 37 Galactic and 84 LMC Cepheids
obtained by \citet{monson12} and \citet{vicky11}, respectively, using {\it
 Spitzer} and IRAC channels 1 \& 2. The variables range in period from 4 to 70
days for the Milky Way and 6 to 140 days for the LMC, and were observed at 24 phase
points from 2009 to 2011.

\section{Fourier Analysis of Galactic and LMC Cepheids}
\label{sec:fparams}

\begin{table*}
\begin{center}
\caption{Fourier parameters obtained using a sine series Fourier fit to the Galactic Cepheid light curves in multi-bands.}
\label{table:fparams_mw}
\scalebox{0.97}{
\begin{tabular}{cccccccccccc}
\hline
\hline
Star ID&  $\log(P)$& $t_{0}$& \emph{N}& $S_{k}(V)$& $A_{c}(V)$& $A_{1}(V)$& $\phi_{1}(V)$& R$_{21}(V)$& R$_{31}(V)$& $\phi_{21}(V)$& $\phi_{31}(V)$ \\
&&(JD)&&&&$\sigma_{A_{1}(V)}$& $\sigma_{\phi_{1}(V)}$& $\sigma_{R_{21}(V)}$& $\sigma_{R_{31}(V)}$& $\sigma_{\phi_{21}(V)}$& $\sigma_{\phi_{31}(V)}$\\
\hline
\hline
AA GEM& 1.05317&       2450327.50 &4& 1.15054& 1.29358& 0.28474& 4.54609& 0.04594& 0.14852 & 1.15955 & 3.44948\\ && & & & & 0.00018& 0.00062& 0.00063 & 0.00064 & 0.01319 & 0.00463 \\
AA MON& 0.59529&       2449805.60 &4& 3.44444& 1.51889& 0.28038& 4.04757& 0.39447& 0.22337 & 2.56312 & 5.53655\\ && & & & & 0.00091& 0.00259& 0.00158 & 0.00326 & 0.01115 & 0.01508 \\

\hline
\hline
Star ID& $\log(P)$& $t_{0}$& \emph{N}& $S_{k}(I)$& $A_{c}(I)$& $A_{1}(I)$& $\phi_{1}(I)$& R$_{21}(I)$& R$_{31}(I)$& $\phi_{21}(I)$& $\phi_{31}(I)$ \\
&&(JD)&&&&$\sigma_{A_{1}(I)}$& $\sigma_{\phi_{1}(I)}$& $\sigma_{R_{21}(I)}$& $\sigma_{R_{31}(I)}$& $\sigma_{\phi_{21}(I)}$& $\sigma_{\phi_{31}(I)}$\\
\hline
\hline
AA MON& 0.59529&       2449809.60 &4& 1.98507& 1.37530& 0.16641& 3.92683& 0.42473& 0.25341 & 3.18826 & 6.21661\\ && & & & & 0.00092& 0.00427& 0.00311 & 0.00559 & 0.01736 & 0.02306 \\
AA SER& 1.23404&       2451255.60 &4& 1.46914& 1.29885& 0.25341& 4.37060& 0.20465& 0.12324 & 2.91294 & 5.09847\\ && & & & & 0.00015& 0.00062& 0.00064 & 0.00060 & 0.00313 & 0.00538 \\

\hline
\hline
Star ID& $\log(P)$& $t_{0}$& \emph{N}& $S_{k}(J)$& $A_{c}(J)$& $A_{1}(J)$& $\phi_{1}(J)$& R$_{21}(J)$& R$_{31}(J)$& $\phi_{21}(J)$& $\phi_{31}(J)$ \\
&&(JD)&&&&$\sigma_{A_{1}(J)}$& $\sigma_{\phi_{1}(J)}$& $\sigma_{R_{21}(J)}$& $\sigma_{R_{31}(J)}$& $\sigma_{\phi_{21}(J)}$& $\sigma_{\phi_{31}(J)}$\\
\hline
\hline
AA GEM& 1.05320&       2454487.90 &4& 1.30415& 0.93050& 0.09333& 4.35333& 0.11786& 0.11175 & 3.08027 & 5.07397\\ && & & & & 0.00420& 0.04400& 0.04489 & 0.04571 & 0.39112 & 0.41752 \\
AA MON& 0.59530&       2454523.80 &5& 1.73224& 0.62075& 0.13759& 3.97621& 0.30773& 0.09935 & 3.68045 & 0.61230\\ && & & & & 0.01007& 0.07183& 0.04312 & 0.06271 & 0.28293 & 0.61763 \\

\hline
\hline
Star ID& $\log(P)$& $t_{0}$& \emph{N}& $S_{k}(H)$& $A_{c}(H)$& $A_{1}(H)$& $\phi_{1}(H)$& R$_{21}(H)$& R$_{31}(H)$& $\phi_{21}(H)$& $\phi_{31}(H)$ \\
&&(JD)&&&&$\sigma_{A_{1}(H)}$& $\sigma_{\phi_{1}(H)}$& $\sigma_{R_{21}(H)}$& $\sigma_{R_{31}(H)}$& $\sigma_{\phi_{21}(H)}$& $\sigma_{\phi_{31}(H)}$\\
\hline
\hline
AA GEM& 1.05320&       2454579.70 &4& 1.28833& 0.81818& 0.10383& 4.43851& 0.11384& 0.02841 & 4.07990 & 4.47053\\ && & & & & 0.00385& 0.03727& 0.03741 & 0.03854 & 0.33727 & 1.28316 \\
AA MON& 0.59530&       2454516.80 &4& 0.75439& 0.32450& 0.10127& 5.12493& 0.34107& 0.25289 & 4.43058 & 2.00580\\ && & & & & 0.00682& 0.06674& 0.05294 & 0.06307 & 0.25083 & 0.33918 \\

\hline
\hline
Star ID& $\log(P)$& $t_{0}$& \emph{N}& $S_{k}(K_{s})$& $A_{c}(K_{s})$& $A_{1}(K_{s})$& $\phi_{1}(K_{s})$& R$_{21}(K_{s})$& R$_{31}(K_{s})$& $\phi_{21}(K_{s})$& $\phi_{31}(K_{s})$ \\
&&(JD)&&&&$\sigma_{A_{1}(K_{s})}$& $\sigma_{\phi_{1}(K_{s})}$& $\sigma_{R_{21}(K_{s})}$& $\sigma_{R_{31}(K_{s})}$& $\sigma_{\phi_{21}(K_{s})}$& $\sigma_{\phi_{31}(K_{s})}$\\
\hline
\hline
AA GEM& 1.05320&       2454579.70 &4& 0.79533& 0.72712& 0.11488& 0.12856& 0.12718& 0.03795 & 5.01138 & 3.95453\\ && & & & & 0.00593& 0.05171& 0.05281 & 0.05453 & 0.41751 & 1.31282 \\
AA MON& 0.59530&       2454492.80 &5& 1.55102& 0.35318& 0.09332& 5.16933& 0.39145& 0.24175 & 4.52406 & 2.11436\\ && & & & & 0.01086& 0.14467& 0.08171 & 0.11754 & 0.45003 & 0.60624 \\

\hline
\hline
Star ID& $\log(P)$& $t_{0}$& \emph{N}& $S_{k}(3.6)$& $A_{c}(3.6)$& $A_{1}(3.6)$& $\phi_{1}(3.6)$& R$_{21}(3.6)$& R$_{31}(3.6)$& $\phi_{21}(3.6)$& $\phi_{31}(3.6)$ \\
&&(JD)&&&&$\sigma_{A_{1}(3.6)}$& $\sigma_{\phi_{1}(3.6)}$& $\sigma_{R_{21}(3.6)}$& $\sigma_{R_{31}(3.6)}$& $\sigma_{\phi_{21}(3.6)}$& $\sigma_{\phi_{31}(3.6)}$\\
\hline
\hline
BETA DOR& 0.99300&      2455185.614 &4& 0.66113& 0.91939& 0.09493& 4.81700& 0.12778& 0.02570 & 5.97377 & 0.57282\\ && & & & & 0.00516& 0.04490& 0.05271 & 0.05111 & 0.41966 & 1.97951 \\
CD CYG& 1.23200&      2455370.893 &4& 1.25734& 0.63132& 0.18164& 4.86587& 0.17023& 0.06485 & 4.51443 & 2.56212\\ && & & & & 0.00368& 0.01584& 0.01920 & 0.01876 & 0.10726 & 0.26134 \\

\hline
\hline
Star ID& $\log(P)$& $t_{0}$& \emph{N}& $S_{k}(4.5)$& $A_{c}(4.5)$& $A_{1}(4.5)$& $\phi_{1}(4.5)$& R$_{21}(4.5)$& R$_{31}(4.5)$& $\phi_{21}(4.5)$& $\phi_{31}(4.5)$ \\
&&(JD)&&&&$\sigma_{A_{1}(4.5)}$& $\sigma_{\phi_{1}(4.5)}$& $\sigma_{R_{21}(4.5)}$& $\sigma_{R_{31}(4.5)}$& $\sigma_{\phi_{21}(4.5)}$& $\sigma_{\phi_{31}(4.5)}$\\
\hline
\hline
BETA DOR& 0.99300&      2455176.550 &4& 0.66667& 0.98020& 0.09632& 5.05796& 0.08150& 0.06209 & 0.02167 & 5.38938\\ && & & & & 0.00296& 0.02814& 0.02721 & 0.02654 & 0.40379 & 0.52805 \\
CD CYG& 1.23200&      2455361.685 &4& 1.31481& 0.71821& 0.18603& 4.46042& 0.16540& 0.02301 & 4.11389 & 1.83127\\ && & & & & 0.00160& 0.00924& 0.00893 & 0.01038 & 0.06042 & 0.35669 \\
\hline
\hline
\end{tabular}}
\end{center}
{\footnotesize \textbf{Notes}: This table is available entirely in a machine-readable form in the online journal.}
\end{table*}

\begin{table*}
\begin{center}
\caption{Fourier parameters obtained using a sine series Fourier fit to the LMC Cepheid light curves in multi-bands. 
The Star ID's are from corresponding catalogues listed in Table~\ref{tab:data_table} for each band.}
\label{table:fparams_lmc}
\scalebox{0.98}{
\begin{tabular}{cccccccccccc}
\hline
\hline
Star ID& $\log(P)$& $t_{0}$& \emph{N}& $S_{k}(V)$& $A_{c}(V)$& $A_{1}(V)$& $\phi_{1}(V)$& R$_{21}(V)$& R$_{31}(V)$& $\phi_{21}(V)$& $\phi_{31}(V)$ \\
&&(JD)&&&&$\sigma_{A_{1}(V)}$& $\sigma_{\phi_{1}(V)}$& $\sigma_{R_{21}(V)}$& $\sigma_{R_{31}(V)}$& $\sigma_{\phi_{21}(V)}$& $\sigma_{\phi_{31}(V)}$\\
\hline
\hline
0002& 0.49389&      2452171.239 &4& 2.44828& 1.45700& 0.19118& 4.19231& 0.33591& 0.12517 & 2.66724 & 5.50564\\ && & & & & 0.00175& 0.00844& 0.00951 & 0.00860 & 0.03024 & 0.07721 \\
0005& 0.74912&      2452171.781 &4& 2.95257& 1.33100& 0.34500& 3.97419& 0.45217& 0.16609 & 3.03606 & 5.91411\\ && & & & & 0.00104& 0.00303& 0.00328 & 0.00286 & 0.00910 & 0.02181 \\

\hline
\hline
Star ID& $\log(P)$& $t_{0}$& \emph{N}& $S_{k}(I)$& $A_{c}(I)$& $A_{1}(I)$& $\phi_{1}(I)$& R$_{21}(I)$& R$_{31}(I)$& $\phi_{21}(I)$& $\phi_{31}(I)$ \\
&&(JD)&&&&$\sigma_{A_{1}(I)}$& $\sigma_{\phi_{1}(I)}$& $\sigma_{R_{21}(I)}$& $\sigma_{R_{31}(I)}$& $\sigma_{\phi_{21}(I)}$& $\sigma_{\phi_{31}(I)}$\\
\hline
\hline
0002& 0.49389&      2452171.239 &4& 2.17460& 1.00803& 0.11220& 4.11196& 0.29546& 0.10232 & 3.14967 & 6.16518\\ && & & & & 0.00051& 0.00470& 0.00474 & 0.00457 & 0.01856 & 0.04814 \\
0005& 0.74912&      2452171.781 &8& 3.14938& 0.77305& 0.20866& 3.84936& 0.43113& 0.16673 & 3.40130 & 0.25359\\ && & & & & 0.00044& 0.00222& 0.00234 & 0.00218 & 0.00670 & 0.01464 \\

\hline
\hline
Star ID& $\log(P)$& $t_{0}$& \emph{N}& $S_{k}(J)$& $A_{c}(J)$& $A_{1}(J)$& $\phi_{1}(J)$& R$_{21}(J)$& R$_{31}(J)$& $\phi_{21}(J)$& $\phi_{31}(J)$ \\
&&(JD)&&&&$\sigma_{A_{1}(J)}$& $\sigma_{\phi_{1}(J)}$& $\sigma_{R_{21}(J)}$& $\sigma_{R_{31}(J)}$& $\sigma_{\phi_{21}(J)}$& $\sigma_{\phi_{31}(J)}$\\
\hline
\hline
0504& 1.15812&      2454438.279 &4& 1.51889& 0.76056& 0.24890& 4.33626& 0.13427& 0.11539 & 3.26927 & 1.21061\\ && & & & & 0.01604& 0.07310& 0.10346 & 0.09234 & 0.24731 & 0.53270 \\
0519& 0.22527&      2454422.182 &6& 0.92678& 0.68350& 0.22571& 3.81431& 0.28187& 0.16313 & 3.91709 & 1.58920\\ && & & & & 0.01811& 0.05389& 0.09476 & 0.07209 & 0.25880 & 0.62753 \\

\hline
\hline
Star ID& $\log(P)$& $t_{0}$& \emph{N}& $S_{k}(H)$& $A_{c}(H)$& $A_{1}(H)$& $\phi_{1}(H)$& R$_{21}(H)$& R$_{31}(H)$& $\phi_{21}(H)$& $\phi_{31}(H)$ \\
&&(JD)&&&&$\sigma_{A_{1}(H)}$& $\sigma_{\phi_{1}(H)}$& $\sigma_{R_{21}(H)}$& $\sigma_{R_{31}(H)}$& $\sigma_{\phi_{21}(H)}$& $\sigma_{\phi_{31}(H)}$\\
\hline
\hline
0494& 0.43568&      2454046.364 &5& 0.17509& 0.16279& 0.17578& 6.10606& 1.53163& 1.06559 & 0.32741 & 5.88143\\ && & & & & 0.06525& 0.51883& 0.80457 & 0.59504 & 1.03915 & 1.56241 \\
0539& 0.53844&      2454426.226 &4& 0.77620& 1.19298& 0.21535& 3.32530& 0.34878& 0.35607 & 6.18024 & 2.34390\\ && & & & & 0.05844& 0.13329& 0.19470 & 0.16559 & 0.57905 & 0.48191 \\

\hline
\hline
Star ID& $\log(P)$& $t_{0}$& \emph{N}& $S_{k}(K_{s})$& $A_{c}(K_{s})$& $A_{1}(K_{s})$& $\phi_{1}(K_{s})$& R$_{21}(K_{s})$& R$_{31}(K_{s})$& $\phi_{21}(K_{s})$& $\phi_{31}(K_{s})$ \\
&&(JD)&&&&$\sigma_{A_{1}(K_{s})}$& $\sigma_{\phi_{1}(K_{s})}$& $\sigma_{R_{21}(K_{s})}$& $\sigma_{R_{31}(K_{s})}$& $\sigma_{\phi_{21}(K_{s})}$& $\sigma_{\phi_{31}(K_{s})}$\\
\hline
\hline
0499& 0.94601&      2454107.035 &4& 0.97239& 0.63132& 0.10926& 4.74642& 0.17893& 0.13381 & 4.95483 & 2.32688\\ && & & & & 0.01208& 0.15320& 0.13889 & 0.13772 & 1.02475 & 1.13511 \\
0504& 1.15812&      2454426.228 &6& 0.67504& 3.34783& 0.14147& 4.55604& 0.29816& 0.43543 & 5.49006 & 3.30983\\ && & & & & 0.02404& 0.71908& 0.22381 & 0.51429 & 3.26315 & 2.84190 \\

\hline
\hline
Star ID& $\log(P)$& $t_{0}$& \emph{N}& $S_{k}(3.6)$& $A_{c}(3.6)$& $A_{1}(3.6)$& $\phi_{1}(3.6)$& R$_{21}(3.6)$& R$_{31}(3.6)$& $\phi_{21}(3.6)$& $\phi_{31}(3.6)$ \\
&&(JD)&&&&$\sigma_{A_{1}(3.6)}$& $\sigma_{\phi_{1}(3.6)}$& $\sigma_{R_{21}(3.6)}$& $\sigma_{R_{31}(3.6)}$& $\sigma_{\phi_{21}(3.6)}$& $\sigma_{\phi_{31}(3.6)}$\\
\hline
\hline
HV1002& 1.48390&      2455153.441 &4& 0.91571& 0.45985& 0.20495& 4.66968& 0.22986& 0.12179 & 4.64503 & 3.12137\\ && & & & & 0.00305& 0.01473& 0.01508 & 0.01548 & 0.07240 & 0.12744 \\
HV1003& 1.38630&      2455144.947 &4& 0.95312& 0.47929& 0.15681& 4.82964& 0.21230& 0.14030 & 4.93681 & 2.83612\\ && & & & & 0.00333& 0.02091& 0.02109 & 0.02088 & 0.11005 & 0.16887 \\

\hline
\hline
Star ID& $\log(P)$& $t_{0}$& \emph{N}& $S_{k}(4.5)$& $A_{c}(4.5)$& $A_{1}(4.5)$& $\phi_{1}(4.5)$& R$_{21}(4.5)$& R$_{31}(4.5)$& $\phi_{21}(4.5)$& $\phi_{31}(4.5)$ \\
&&(JD)&&&&$\sigma_{A_{1}(4.5)}$& $\sigma_{\phi_{1}(4.5)}$& $\sigma_{R_{21}(4.5)}$& $\sigma_{R_{31}(4.5)}$& $\sigma_{\phi_{21}(4.5)}$& $\sigma_{\phi_{31}(4.5)}$\\
\hline
\hline
HV1002& 1.48390&      2455152.041 &4& 1.33645& 0.50830& 0.21650& 4.63455& 0.22088& 0.13594 & 4.21154 & 2.92194\\ && & & & & 0.00330& 0.01493& 0.01534 & 0.01465 & 0.07610 & 0.12536 \\
HV1003& 1.38630&      2455142.933 &4& 1.18818& 0.60256& 0.16166& 4.58925& 0.18291& 0.17568 & 4.63619 & 2.19634\\ && & & & & 0.00364& 0.02225& 0.02198 & 0.02219 & 0.13488 & 0.14987 \\

\hline
\hline
\end{tabular}}
\end{center}
{\footnotesize \textbf{Notes}: This table is available entirely in a machine-readable form in the online journal.\\
Star ID's consisting of 4 numbers are based on the OGLE-III catalog.}
\end{table*}

We applied the Fourier decomposition method discussed in 
Section~\ref{sec:fourier} individually to each Galactic and LMC Cepheid light curve,
analyzing each bandpass separately. We implemented equation~(\ref{eq:foufit1})
using the IDL MPCURVEFIT routine, varying the order of the fit in each band from 4 to 8.
The optimum order of fit (\emph{N}) was determined 
using Baart's condition, depending on the residuals for each star
\citep{baart82, deb09}. The resulting Fourier coefficients were used
to calculate Fourier parameters using equation~(\ref{eq:fparams}).
Fourier-fitted light curves for two Galactic and two LMC Cepheids are 
shown in Fig.~\ref{fig:galactic_lc.eps} and \ref{fig:lmc_lc.eps}, respectively.
The Fourier parameters for all variables in all bands are presented in
Tables~\ref{table:fparams_mw} and \ref{table:fparams_lmc} for Galactic and LMC
Cepheids, respectively. In all the figures presented in our paper, the values of the Fourier phase
parameter $\phi_{31}$, obtained using a sine series, were converted into cosine series by adding a value of $\pi$ to 
those given in Table~\ref{table:fparams_mw} and \ref{table:fparams_lmc} \citep{deb14}.

\subsection{Optical Bands}

\begin{figure*}
\begin{center}
\includegraphics[width=1.0\textwidth,keepaspectratio]{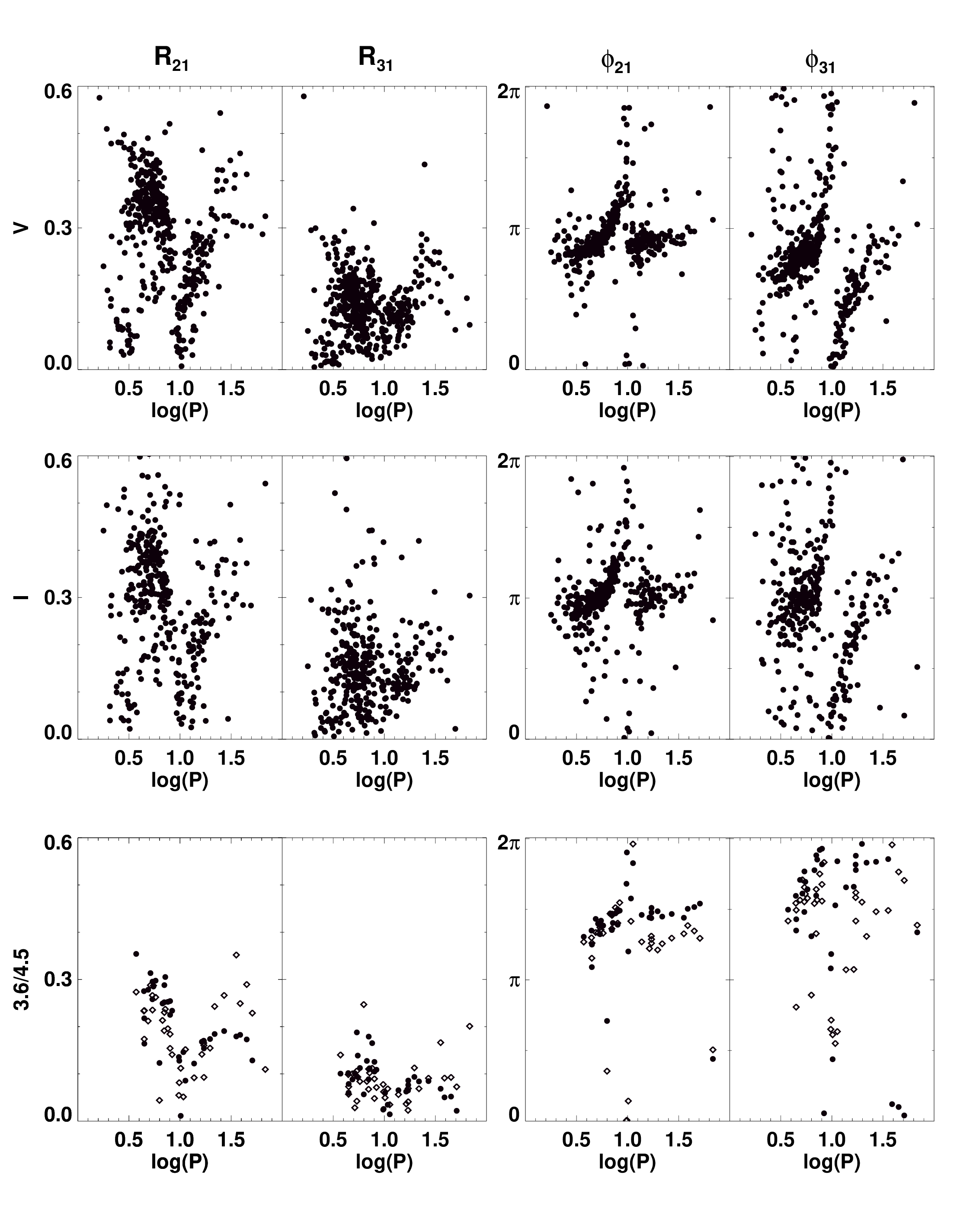}
\caption{Galactic Cepheid Fourier parameters for the V (top row), I (middle row) and
  mid-infrared (bottom row) bands. In the mid-infrared panels, circles and
  diamonds represent 3.6 and 4.5$\mu$m data, respectively. The first two columns
  show Fourier amplitude parameters ($R_{21}$ $\&$ $R_{31}$) while the last two
  represent Fourier phase parameters ($\phi_{21}$ $\&$ $\phi_{31}$).}
\label{fig:mw_fou_lamda_mir.eps}
\end{center}
\end{figure*}

We have determined the Fourier parameters ($R_{21}$, $R_{31}$, $\phi_{21}$ $\&$
$\phi_{31}$) of one of the largest
samples of Galactic Cepheids at optical wavelengths, with 447 objects in $V$ and
351 in $I$. These are shown in the top two rows of
Fig.~\ref{fig:mw_fou_lamda_mir.eps} as a function of $\log(P)$.

The Hertzsprung progression (hereafter HP), indicated by a sharp 
dip or change in the way the Fourier parameters change with period, is clearly observed for $R_{21}$, 
$\phi_{21}$ $\&$ $\phi_{31}$ in the vicinity of $\log(P)=1.0$. 
The center of this HP seems to be located at slightly shorter
period. $R_{31}$ exhibits a flatter minimum that extends over a couple of days in the vicinity 
of $\log(P)=1.0$. The uncertainties in the Fourier parameters are very small,
given the fairly good phase coverage and number of data points in the light curves.

\begin{figure*}
\begin{center}
\includegraphics[width=1.0\textwidth,keepaspectratio]{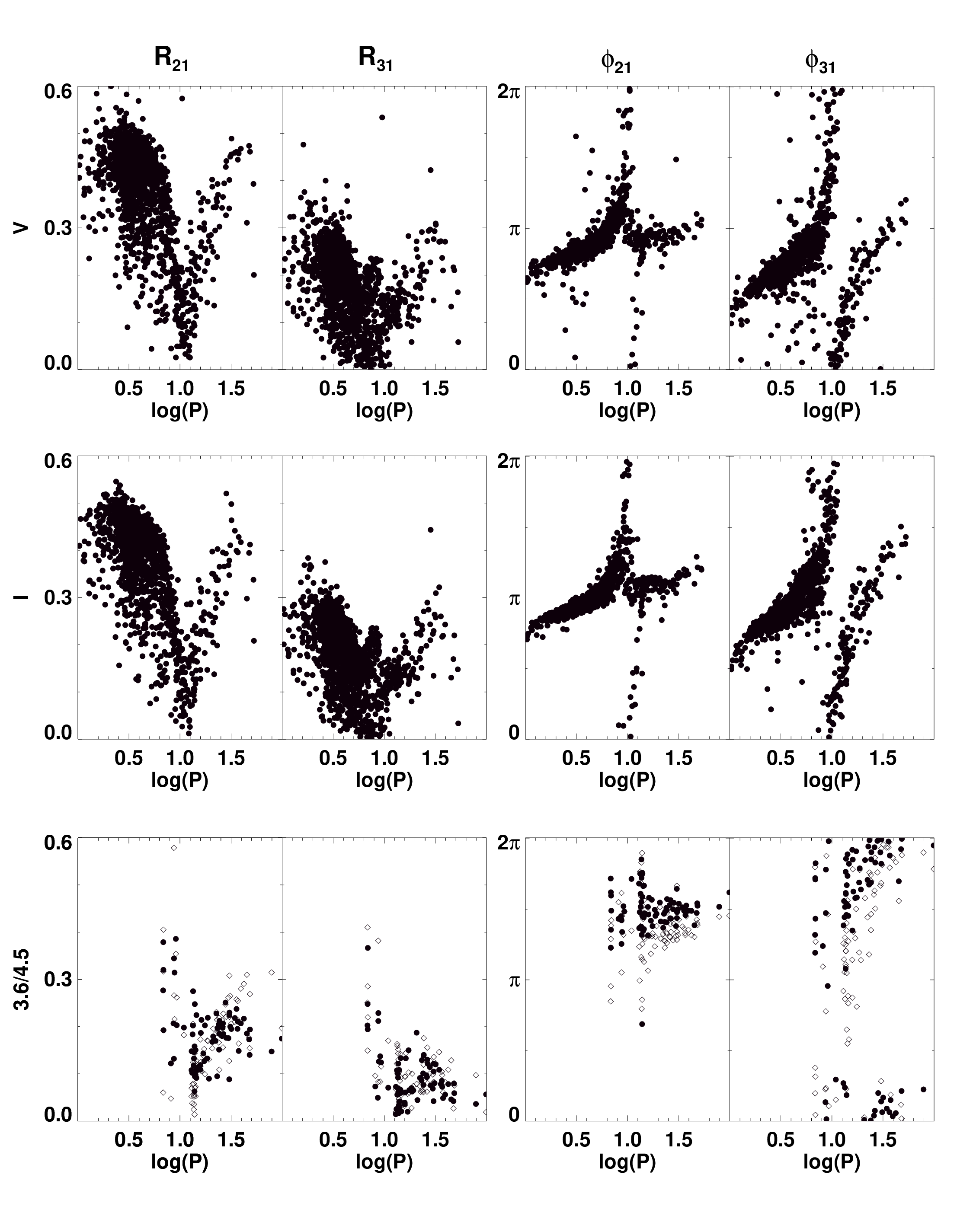}
\caption{LMC Cepheid Fourier parameters, presented in the same manner as Fig.~\ref{fig:mw_fou_lamda_mir.eps}.}
\label{fig:lmc_fou_lamda_mir.eps}
\end{center}
\end{figure*}

The top two rows of Fig.~\ref{fig:lmc_fou_lamda_mir.eps} show the corresponding analysis for
OGLE-III LMC Cepheids in $V$- and $I$-bands. These light curves have excellent 
sampling and full phase coverage, thereby yielding very well determined
Fourier parameters that show clear patterns as discussed by 
\citet{sosz2008} $\&$ \citet{ulac13} in their data release papers.

\begin{figure*}
\begin{center}
\includegraphics[width=1.0\textwidth,keepaspectratio]{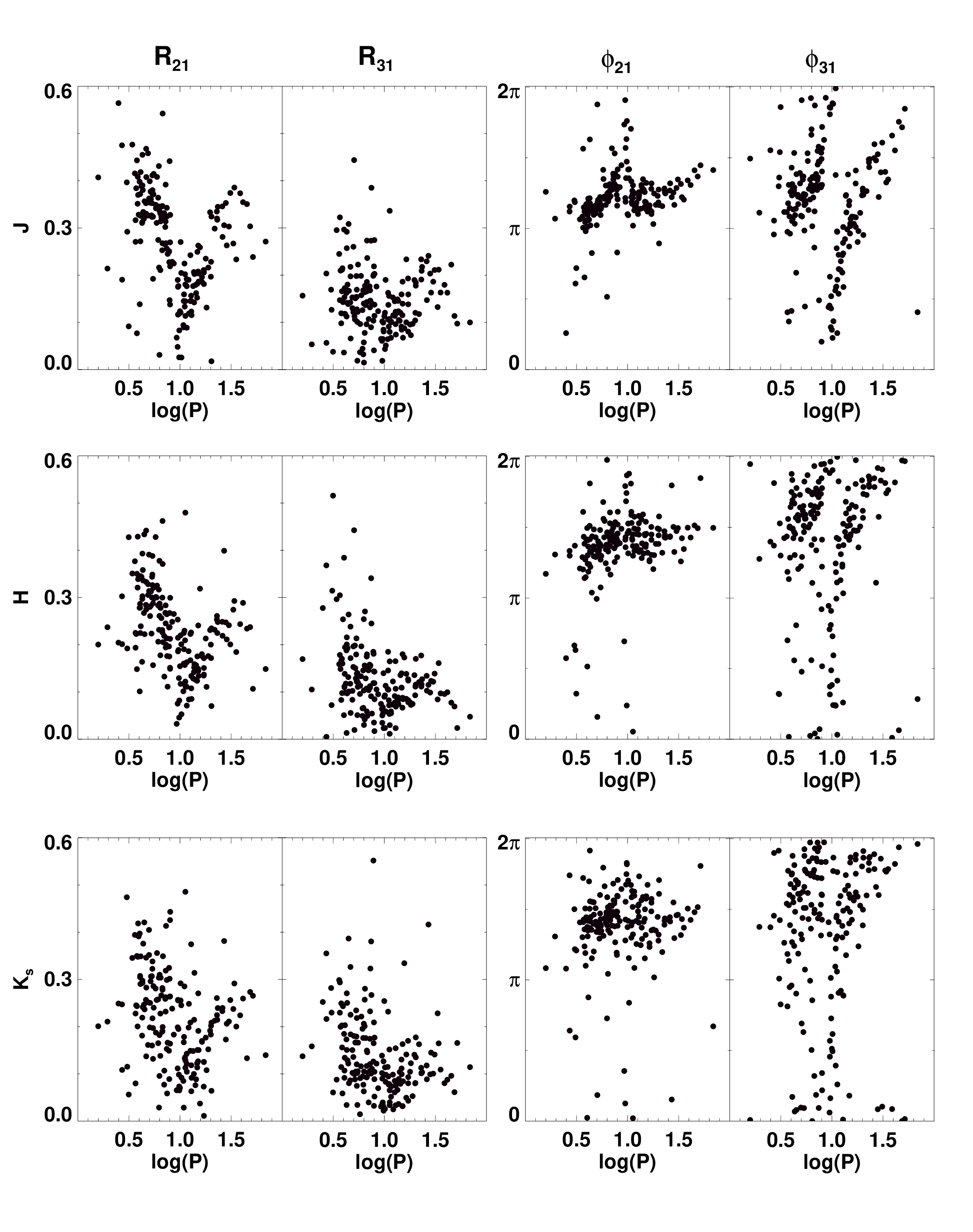}
\caption{Galactic Cepheid Fourier parameters for the J (top row), H (middle row) and
  K$_s$ (bottom row) bands, arranged in the column order as Fig.~\ref{fig:mw_fou_lamda_mir.eps}.}
\label{fig:mw_fou_lamda_nir.eps}
\end{center}
\end{figure*}

\begin{figure*}
\begin{center}
\includegraphics[width=1.0\textwidth,keepaspectratio]{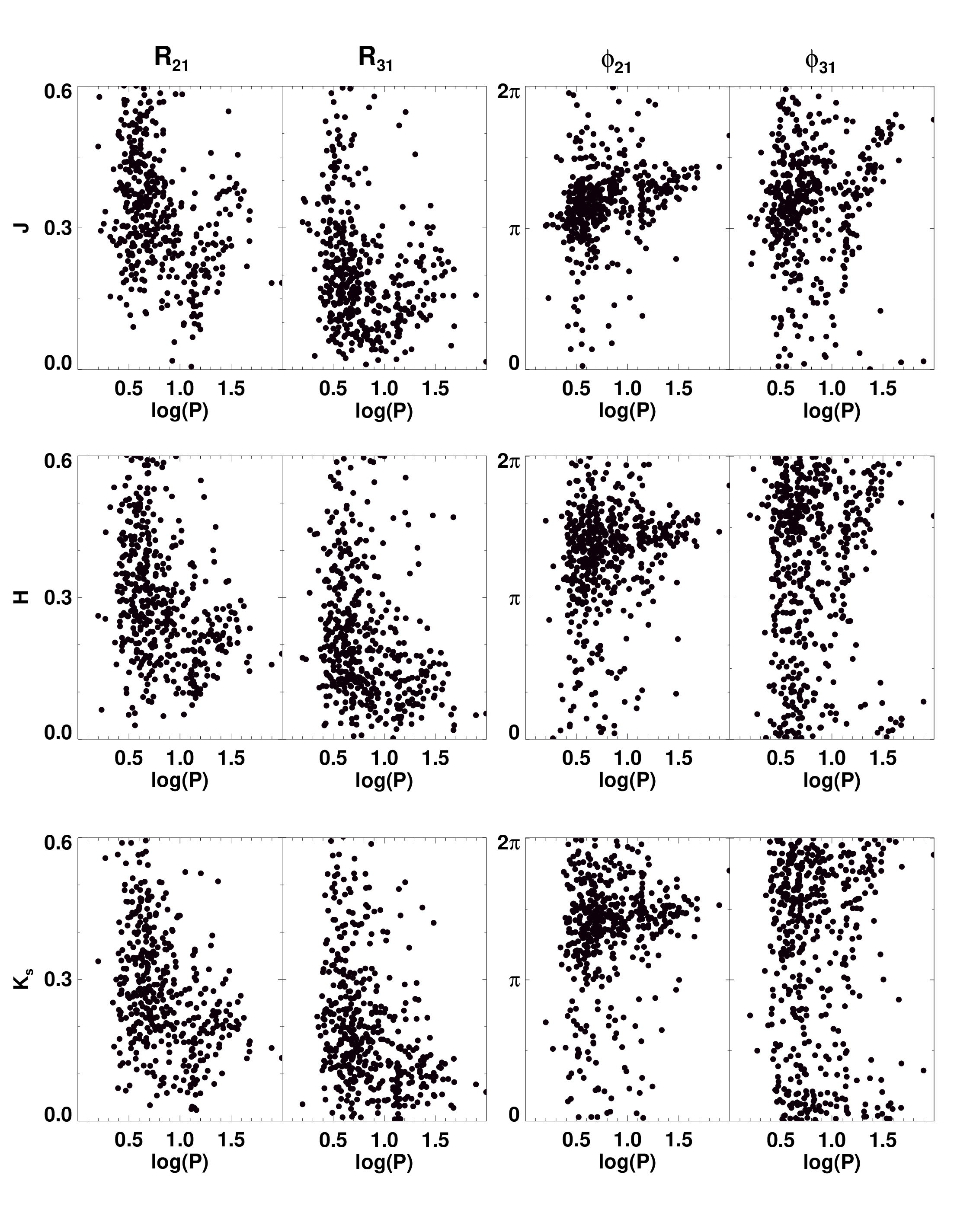}
\caption{LMC Cepheid Fourier parameters, presented in the same manner as Fig.~\ref{fig:mw_fou_lamda_nir.eps}.}
\label{fig:lmc_fou_lamda_nir.eps}
\end{center}
\end{figure*}

\begin{figure}
\begin{center}
\includegraphics[width=0.5\textwidth,keepaspectratio]{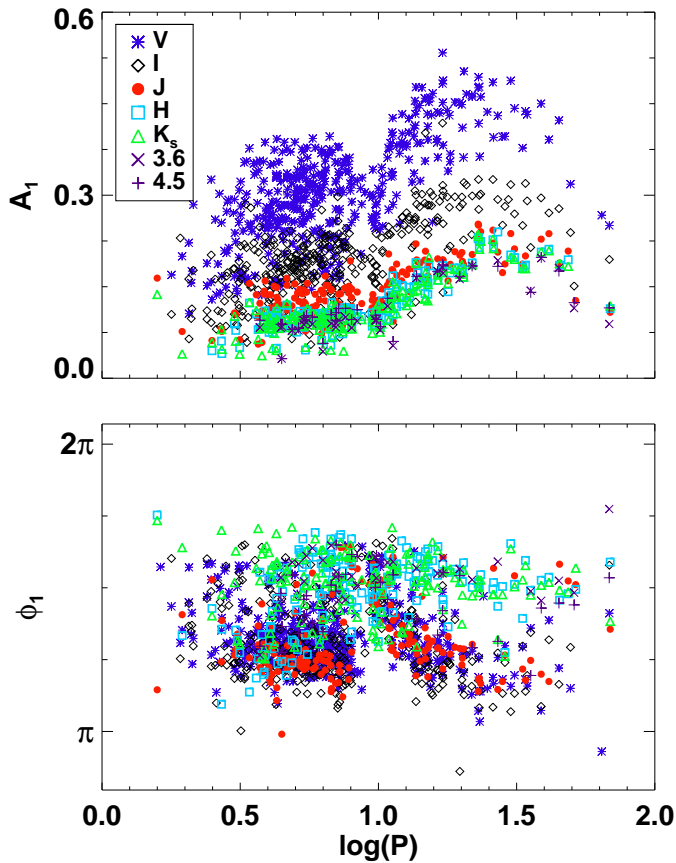}
\end{center}
\caption{ Variation of Fourier amplitude ($A_{1}$) and phase ($\phi_{1}$) coefficients for Galactic Cepheids in multiple bands.} 
\label{fig:a1_phi1_mw.eps}
\end{figure}

\begin{figure}
\begin{center}
\includegraphics[width=0.5\textwidth,keepaspectratio]{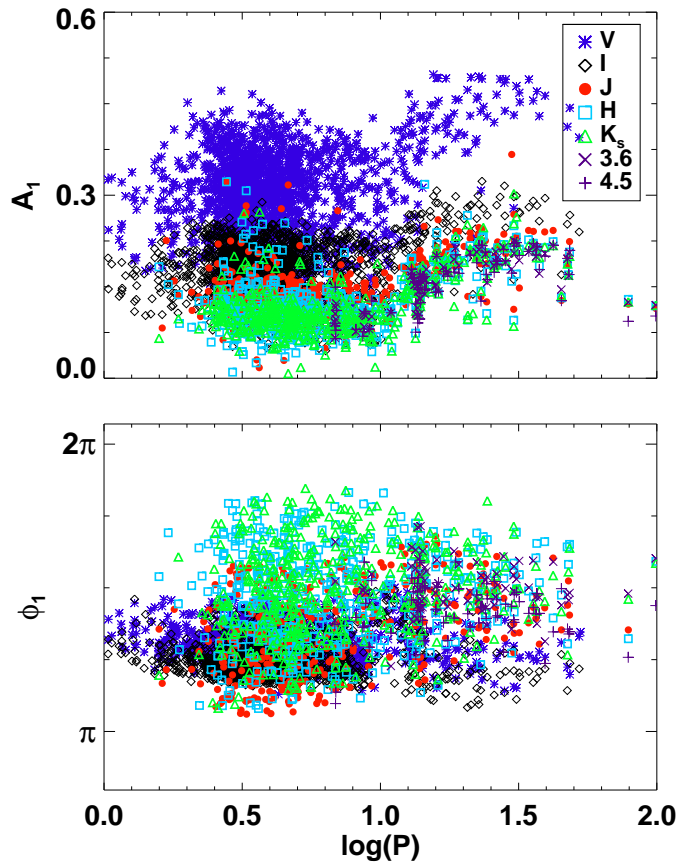}
\end{center}
\caption{ Same as Fig.~\ref{fig:a1_phi1_mw.eps}, but for LMC Cepheids.} 
\label{fig:a1_phi1_lmc.eps}
\end{figure}

\subsection{Near-Infrared Bands}

Fig.~\ref{fig:mw_fou_lamda_nir.eps} presents the Fourier parameters for 186
Galactic Cepheids that presently have full phase coverage. The HP is clearly observed for all parameters in the vicinity of 
$\log(P)=1.0$. We also noticed a more gradual increase of 
$\phi_{31}$ for $\log(P)>1.0$ with increasing wavelength.
Since the near-infrared data have a lower number of epochs and poorer phase
coverage, the errors in the parameters are larger than
at optical bands. The $J$-band parameters are the best determined ones, with
increasing scatter at $H$ and $K_{\rm s}$.

Fig.~\ref{fig:lmc_fou_lamda_nir.eps} presents the corresponding results for the near-infrared
LMC data, including the first-ever Fourier analysis of the light curves from 
\citet{macri14}. Since these light curves are not as well sampled
as their optical counterparts, we observe more scatter in the Fourier
parameters. The better-sampled light curves from \citet{persson04}, which
predominantly cover variables with $\log(P)>1.0$, enable us to clearly see the HP 
in the $J$-band Fourier parameters, while the $H$ and $K_{s}$ panels exhibit greater scatter.

\subsection{Mid-Infrared Bands}

The mid-infrared Fourier parameters for Galactic Cepheids are plotted in the bottom row of 
Fig.~\ref{fig:mw_fou_lamda_mir.eps}, showing for the first time the
variation of light curve structure as a function of period at these wavelengths.
Even with a smaller number of Cepheids in the sample, the HP is clearly visible for all Fourier parameters. The value of $R_{21}$
displays an abrupt rise from $\log(P)=1.0$ to a maximum value of $\log(P)=1.5$ at
4.5-$\mu$m, which is not seen at 3.6-$\mu$m. Since these light curves have equal
phase spacing and the same number of data points, the errors in the parameters are smaller than in their near-infrared counterparts.

The corresponding parameters for LMC Cepheids are displayed in the bottom row of 
Fig.~\ref{fig:lmc_fou_lamda_mir.eps}. We observe similar patterns to those
exhibited by the Galactic variables.

\section{Comparison of Fourier Parameters}
\label{sec:comp_fparams}

\begin{figure*}
\begin{center}
\includegraphics[width=1.0\textwidth,keepaspectratio]{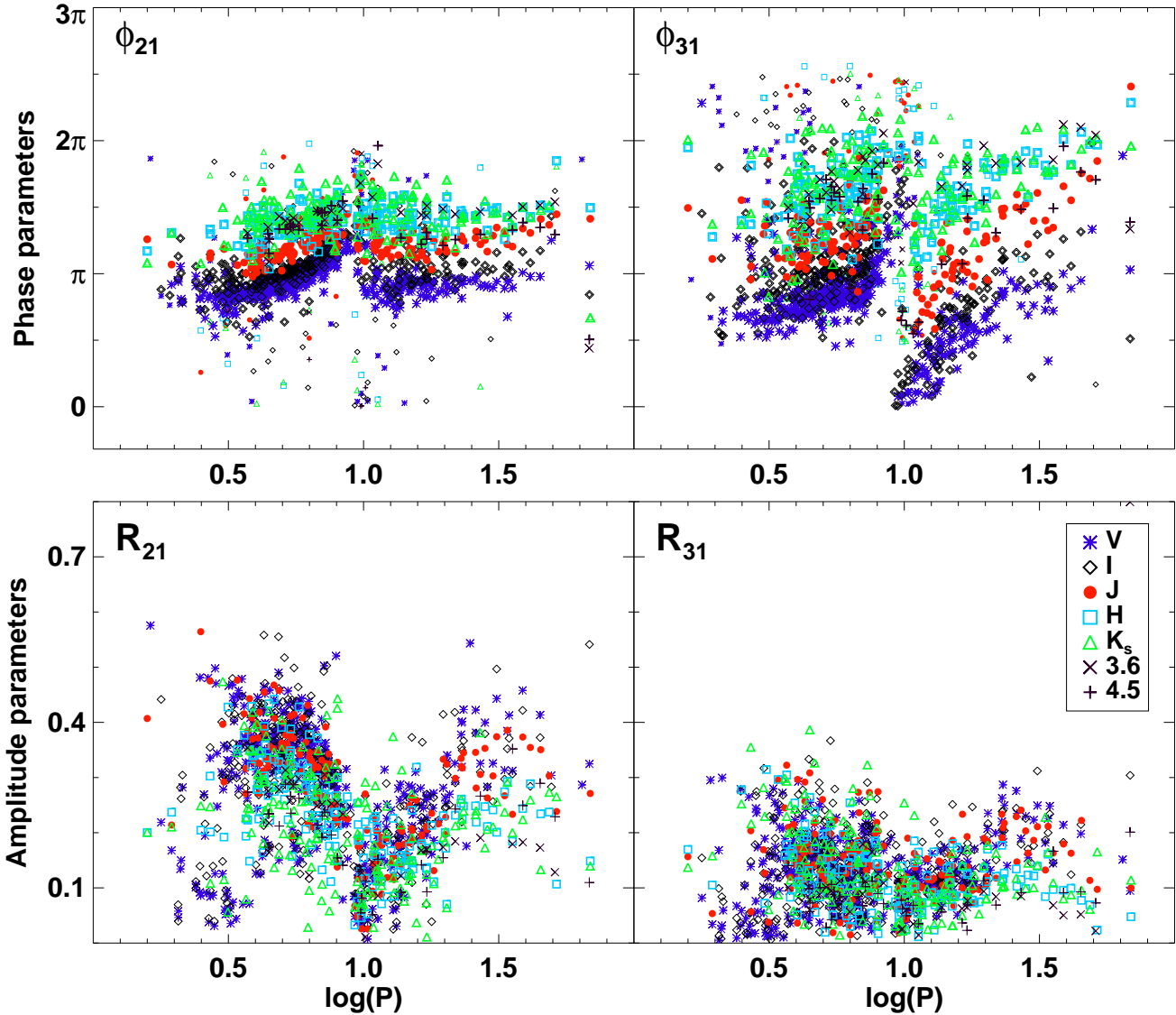}
\end{center}
\caption{ Fourier parameters plotted against $\log(P)$ for Galactic Cepheids, in multiple bands. Some phase parameters have
been shifted by $2\pi$ for plotting purposes. The outliers in phase parameters for each band are shown using smaller symbols.}
\label{fig:fou_multi_mw.eps}
\end{figure*}

\begin{figure*}
\begin{center}
\includegraphics[width=1.0\textwidth,keepaspectratio]{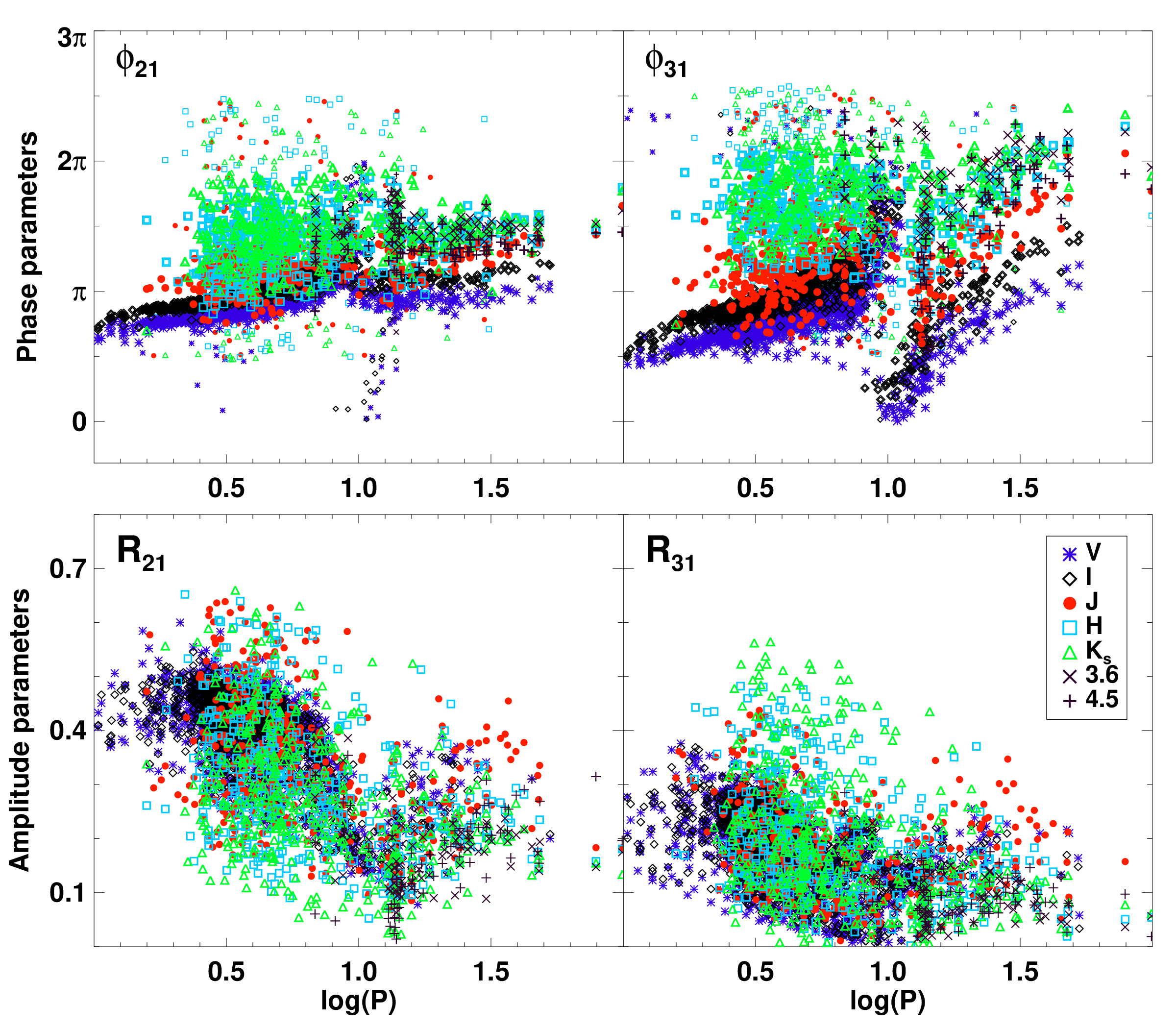}
\end{center}
\caption{Same as Fig.~\ref{fig:fou_multi_mw.eps}, but for LMC Cepheids.}
\label{fig:fou_multi_lmc.eps}
\end{figure*}

\subsection{Fourier amplitude and phase coefficient}

We discuss the variation of the first harmonic of amplitude ($A_{1}$) and first Fourier phase coefficient ($\phi_{1}$) 
with period and wavelength. The plots are shown in Fig.~\ref{fig:a1_phi1_mw.eps} and \ref{fig:a1_phi1_lmc.eps},
for Galactic and LMC Cepheids respectively. We removed the $2\sigma$ outliers in these plots to make the 
progression visible. At a given period, we observe a decrease in the value of $A_{1}$ and an increase in $\phi_{1}$ with 
increasing wavelength for both Galaxy and LMC. For $1.0 < \log(P) < 1.5$, we find a greater increase in the value of $A_{1}$ 
and decrease in $\phi_{1}$ at optical bands as compared to their infrared counterparts. Also, for $\log(P) > 1.3$, both 
coefficients show a very small variation and nearly a flat curve at infrared bands. A larger scatter in the value of 
$\phi_{1}$ is observed for infrared bands.

We note that the variation of light curve amplitude is essentially similar to $A_{1}$ but the
variation of phase of maximum light (corresponding to $t_{0}$) is not necessarily same as $\phi_{1}$.
As the Infrared band light curves have larger phase gaps and $\phi_{1}$ depends on the value of $t_{0}$, 
we expect a greater scatter in these bands as observed in Fig.~\ref{fig:a1_phi1_mw.eps} and \ref{fig:a1_phi1_lmc.eps}. 
For example, in Fig.~\ref{fig:galactic_lc.eps} the $J$-band light curve for $T~MON$ show a flatter maxima, which 
also causes an uncertainty in the determination of exact phase corresponding to maximum light.
The phase difference at maximum light for Cepheids in the Galaxy in multiple bands is
discussed in \citet{madore91} while $\Delta \phi_{max}(I~vs.~JHK_{s})$ for first results of 
LMC Cepheids used in our analysis is discussed in \citet{macri14}. This phase difference at maximum 
light and its variation with wavelength is related to Period-Color relations at maximum light discussed in \citet{bhardwaj14}.

We also emphasize that the coefficients $\phi_{i}$ are not independent of time translation and this is the reason 
phase parameters (${\phi}_{i1}$) are preferred to study the light curve structure \citep{slee81}. These are perhaps
more easily compared between data sets where the initial epoch of observation is not known. 
The variation of phase parameters with period and wavelength will be discussed in the following subsections.

\subsection{Individual Fourier Parameters}

In order to analyze the variation of Fourier parameters as a function of
wavelength and period, we over-plotted the phase and amplitude parameters using
different symbols and colors for each band, and removed 2$\sigma$ outliers to
make the Hertzsprung progressions more easily
visible. Figs.~\ref{fig:fou_multi_mw.eps} and
\ref{fig:fou_multi_lmc.eps} present the parameters for the Galactic and LMC
samples, respectively. 

We observed a clear trend in the $\phi_{21}$ $\&$
$\phi_{31}$ for both Galactic and LMC Cepheids that become larger with increasing
wavelength at fixed period. For these phase parameters the outliers are shown using smaller symbols for all bands to avoid any
loss of features near the center of the HP. These outliers follow the same trend as all points and the features of the plot are not 
affected by varying the degree of outlier removal.
No clear trend with wavelength at fixed period is seen for the
$R_{21}$ and $R_{31}$. Considering long-period ($\log(P)>1.0$)
variables, the amplitude parameters can be separated into two groups; one for
$VIJ$ and another for the longer wavelengths. Furthermore, the latter bands
exhibit a slight drop after $\log(P)=1.3$~($P=20$~d) while the former ones seem to rise. We also observed
a turnover in amplitude parameters around $\log(P)=1.5$ that varies with
wavelength. Since we have removed 2$\sigma$ outliers in this Figure, the
Galactic data show more clearly that the center of the Hertzsprung progression occurs
slightly before $\log(P)=1.0$ for all parameters while it remains at
$\log(P)=1.0$ for the LMC variables.

\subsection{Mean Fourier Parameters}
 
\begin{figure*}
\begin{center}
\includegraphics[width=0.7\textwidth,keepaspectratio]{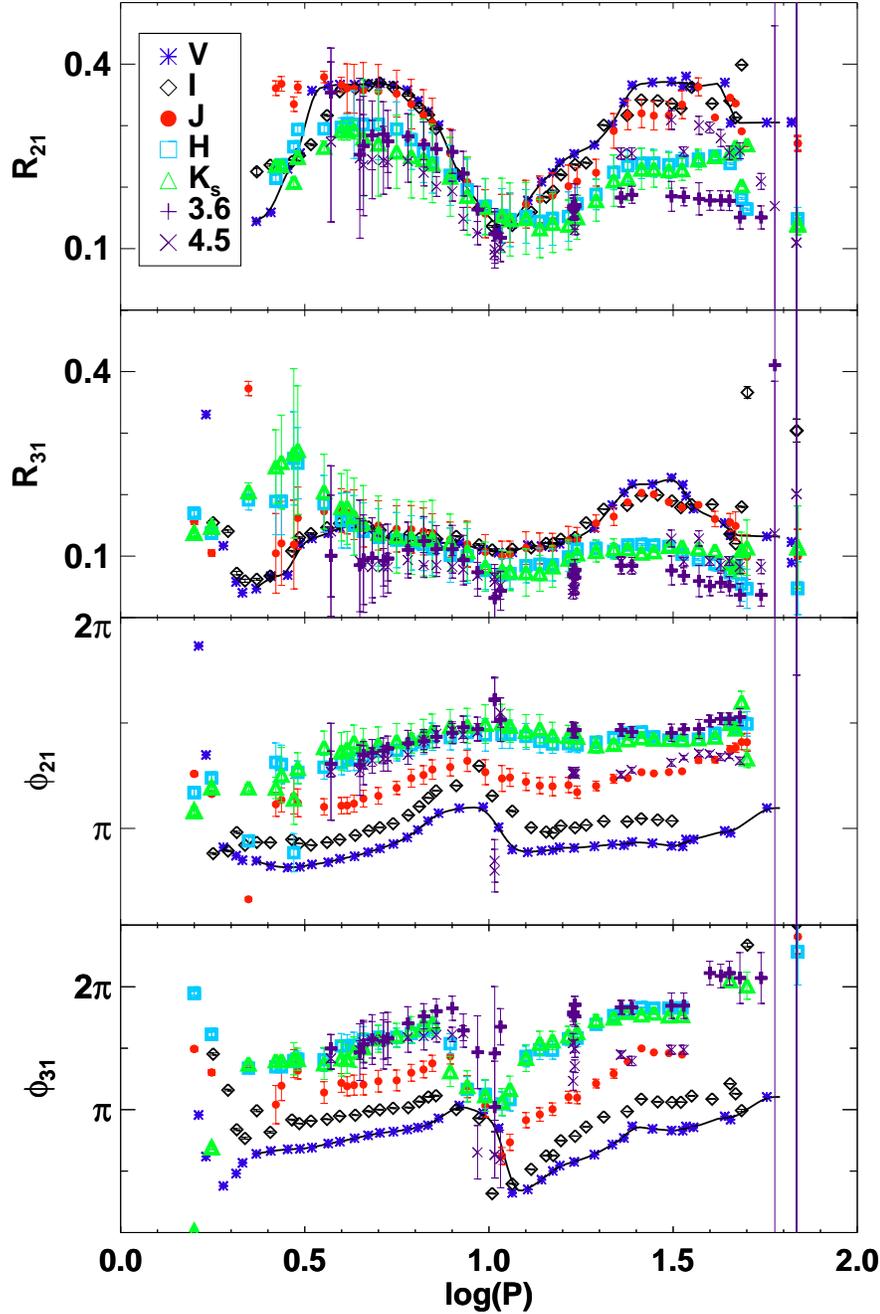}
\end{center}
\caption{ Mean Fourier Parameters for Galactic Cepheids. Some phase parameters have
been shifted by $2\pi$ for plotting purposes. The error bars represent the standard error in the mean values.}

\label{fig:mean_fou_mw.eps}
\end{figure*}

\begin{figure*}
\begin{center}
\includegraphics[width=0.7\textwidth,keepaspectratio]{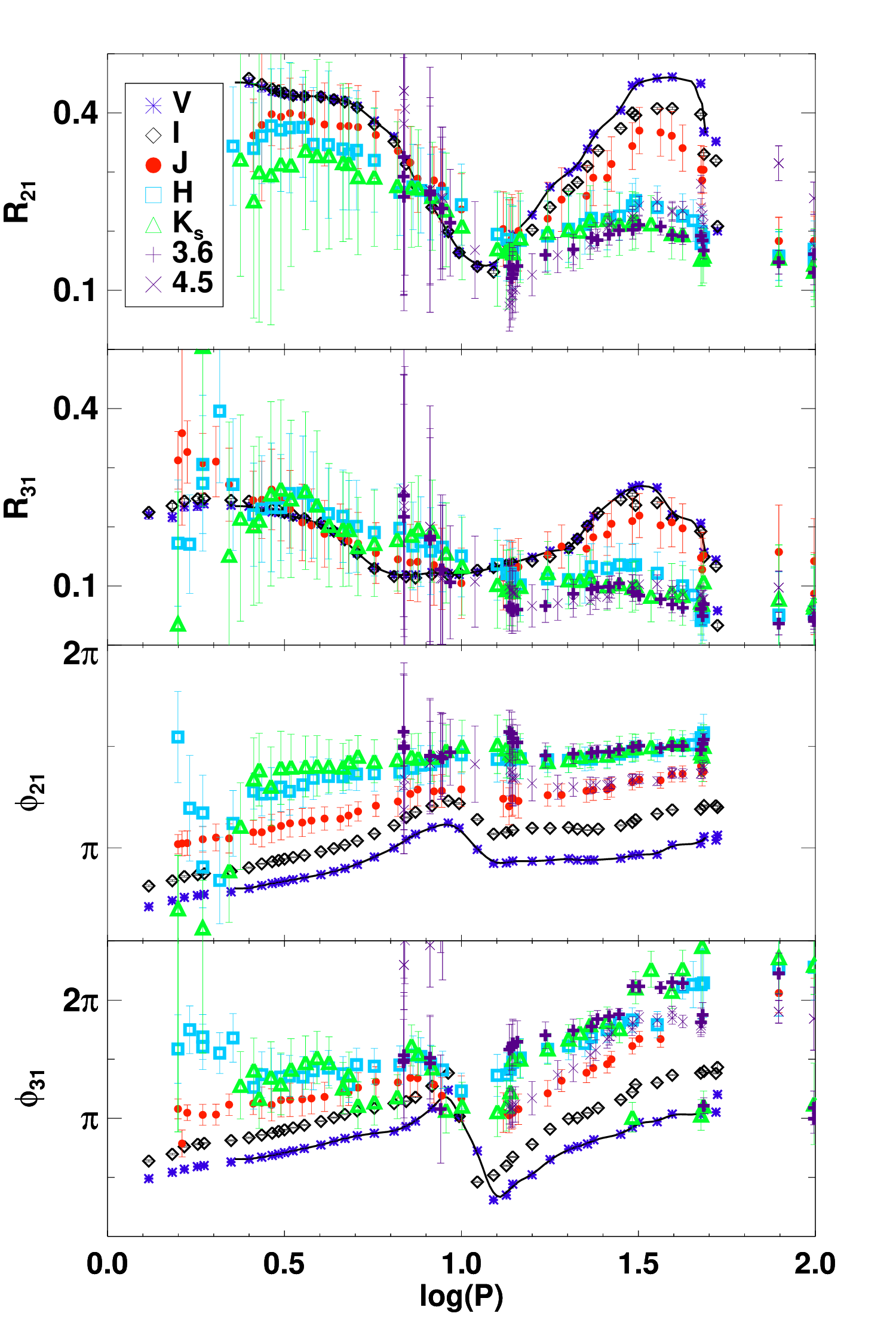}
\end{center}
\caption{ Same as Fig.~\ref{fig:mean_fou_lmc.eps}, but for LMC Cepheids.}
\label{fig:mean_fou_lmc.eps}
\end{figure*}

In order to clearly discern any wavelength dependent variation in Fourier parameters,
we computed sliding mean values with steps of 0.04~dex in $\log (P)$ and a bin width of
0.2~dex. We found these values yielded the least amount of scatter between
consecutive points after significant experimentation with various choices.

Fig.~\ref{fig:mean_fou_mw.eps} shows the result for Galactic Cepheids. The
increase in phase parameters  with increasing
wavelength becomes more clear and distinct.  We also see clearly the decrease in
amplitude parameters with increasing wavelength at a
given period. Both $R_{21}$ $\&$ $R_{31}$ exhibit a sharp rise beyond $\log(P)=0.9$ to a
peak around $\log(P)=1.4$ and a decrease around $\log(P)=1.7$. This behavior is
more pronounced for $VIJ$ than for the redder bands. The minimum is more
pronounced for $R_{21}$, while $R_{31}$ shows a shallower minimum. There is also
a hint of change in the behavior of the parameters for $\log(P)>1.8$. This
may be connected to the properties of ultra long period Cepheids
\citep[ULPCs,][]{ccn13}. Considering the short-period variables, both parameters exhibit
maximum values around $\log(P)=0.6$. The increased scatter for $\log(P)<0.5$
may be due to a combination of a smaller number of stars and contamination by
first overtone pulsators.

\begin{table*}
\begin{center}
\caption{Variation of mean Fourier parameters, determined as a difference in multiple-bands in each period bin of $\log(P)=0.2$.}
\label{table:diff_fparams}
\scalebox{0.96}{
\begin{tabular}{ccccccc}
\hline
\hline
  $\log(P)$ &  $\Delta$R$_{21}(V,I)$  &   $\Delta$R$_{21}(V,K_{s})$ &      $\Delta$R$_{21}(J,K_{s})$ &   $\Delta$R$_{21}(H,K_{s})$ & $\Delta$R$_{21}(3.6,K_{s})$ &  $\Delta$R$_{21}(3.6,4.5)$ \\
\hline
\hline
\multicolumn{7}{|c|}{Galaxy} \\
\hline 
       0.5-0.7       &     0.031$\pm$0.002     &     0.082$\pm$0.064     &     0.081$\pm$0.076     &     0.018$\pm$0.081     &     0.020$\pm$0.097     &     0.047$\pm$0.149     \\
       0.7-0.9       &     0.005$\pm$0.002     &     0.101$\pm$0.059     &     0.071$\pm$0.070     &     0.016$\pm$0.073     &     0.033$\pm$0.070     &     0.045$\pm$0.051     \\
       0.9-1.1       &    -0.003$\pm$0.001     &    -0.001$\pm$0.040     &     0.002$\pm$0.053     &     0.003$\pm$0.057     &     0.004$\pm$0.054     &     0.042$\pm$0.043     \\
       1.1-1.3       &     0.023$\pm$0.001     &     0.080$\pm$0.034     &     0.047$\pm$0.041     &     0.011$\pm$0.042     &     0.020$\pm$0.038     &     0.018$\pm$0.020     \\
       1.3-1.5       &     0.022$\pm$0.000     &     0.127$\pm$0.002     &     0.090$\pm$0.002     &     0.012$\pm$0.002     &    -0.038$\pm$0.015     &    -0.084$\pm$0.018     \\

\hline
\multicolumn{7}{|c|}{ LMC} \\
\hline
       0.5-0.7       &     0.001$\pm$0.005     &     0.105$\pm$0.108     &     0.066$\pm$0.124     &     0.032$\pm$0.130     &    -0.318$\pm$0.108     &     0.000$\pm$0.000     \\
       0.7-0.9       &     0.007$\pm$0.005     &     0.058$\pm$0.098     &     0.047$\pm$0.121     &     0.003$\pm$0.120     &     0.002$\pm$0.191     &    -0.066$\pm$0.356     \\
       0.9-1.1       &     0.003$\pm$0.004     &    -0.057$\pm$0.080     &     0.032$\pm$0.104     &     0.021$\pm$0.105     &    -0.004$\pm$0.145     &    -0.015$\pm$0.170     \\
       1.1-1.3       &     0.026$\pm$0.003     &     0.060$\pm$0.063     &     0.029$\pm$0.078     &     0.011$\pm$0.081     &    -0.039$\pm$0.074     &     0.038$\pm$0.055     \\
       1.3-1.5       &     0.044$\pm$0.003     &     0.163$\pm$0.046     &     0.099$\pm$0.061     &     0.016$\pm$0.059     &    -0.022$\pm$0.049     &    -0.008$\pm$0.028     \\

\hline
\hline
 $\log(P)$ &  $\Delta$R$_{31}(V,I)$  &   $\Delta$R$_{31}(V,K_{s})$ &     $\Delta$R$_{31}(J,K_{s})$ &   $\Delta$R$_{31}(H,K_{s})$ &  $\Delta$R$_{31}(3.6,K_{s})$ &  $\Delta$R$_{31}(3.6,4.5)$ \\
\hline
\hline
\multicolumn{7}{|c|}{ Galaxy} \\
\hline
      0.5-0.7       &    -0.006$\pm$0.001     &    -0.028$\pm$0.061     &    -0.005$\pm$0.072     &    -0.016$\pm$0.076     &    -0.074$\pm$0.097     &    -0.012$\pm$0.125     \\
       0.7-0.9       &    -0.007$\pm$0.002     &    -0.002$\pm$0.060     &     0.011$\pm$0.071     &    -0.012$\pm$0.074     &    -0.021$\pm$0.075     &     0.021$\pm$0.051     \\
       0.9-1.1       &    -0.007$\pm$0.001     &     0.025$\pm$0.040     &     0.022$\pm$0.049     &    -0.006$\pm$0.051     &    -0.021$\pm$0.057     &    -0.003$\pm$0.046     \\
       1.1-1.3       &    -0.005$\pm$0.001     &     0.034$\pm$0.031     &     0.035$\pm$0.039     &     0.007$\pm$0.040     &    -0.017$\pm$0.035     &     0.027$\pm$0.018     \\
       1.3-1.5       &     0.020$\pm$0.000     &     0.099$\pm$0.002     &     0.078$\pm$0.002     &     0.005$\pm$0.002     &    -0.028$\pm$0.013     &    -0.015$\pm$0.016     \\

\hline
\multicolumn{7}{|c|}{ LMC} \\
\hline
     0.5-0.7       &    -0.001$\pm$0.005     &    -0.022$\pm$0.104     &    -0.021$\pm$0.119     &     0.012$\pm$0.130     &    -0.215$\pm$0.104     &     0.000$\pm$0.000     \\
       0.7-0.9       &     0.003$\pm$0.005     &    -0.061$\pm$0.090     &    -0.038$\pm$0.111     &    -0.001$\pm$0.117     &     0.034$\pm$0.210     &    -0.010$\pm$0.292     \\
       0.9-1.1       &    -0.000$\pm$0.004     &    -0.037$\pm$0.085     &    -0.038$\pm$0.108     &    -0.002$\pm$0.112     &    -0.040$\pm$0.152     &     0.015$\pm$0.177     \\
       1.1-1.3       &     0.006$\pm$0.003     &     0.050$\pm$0.064     &     0.034$\pm$0.080     &     0.015$\pm$0.083     &    -0.039$\pm$0.075     &    -0.026$\pm$0.055     \\
       1.3-1.5       &     0.031$\pm$0.003     &     0.141$\pm$0.042     &     0.089$\pm$0.057     &     0.024$\pm$0.056     &    -0.010$\pm$0.046     &     0.008$\pm$0.028     \\

\hline
\hline
 $\log(P)$ &  $\Delta \phi_{21}(V,I)$  &   $\Delta \phi_{21}(V,K_{s})$ &      $\Delta \phi_{21}(J,K_{s})$ &   $\Delta \phi_{21}(H,K_{s})$ & $\Delta \phi_{21}(3.6,K_{s})$ &  $\Delta \phi_{21}(3.6,4.5)$ \\
\hline
\hline
\multicolumn{7}{|c|}{ Galaxy} \\
\hline
       0.5-0.7       &    -0.303$\pm$0.005     &    -1.673$\pm$0.257     &    -0.836$\pm$0.285     &    -0.192$\pm$0.310     &    -0.203$\pm$0.321     &     0.115$\pm$0.422     \\
       0.7-0.9       &    -0.331$\pm$0.006     &    -1.460$\pm$0.250     &    -0.614$\pm$0.283     &    -0.118$\pm$0.311     &    -0.145$\pm$0.310     &     0.080$\pm$0.232     \\
       0.9-1.1       &    -0.499$\pm$0.009     &    -1.466$\pm$0.274     &    -0.709$\pm$0.331     &    -0.140$\pm$0.366     &     0.098$\pm$0.427     &     0.827$\pm$0.392     \\
       1.1-1.3       &    -0.308$\pm$0.004     &    -1.691$\pm$0.194     &    -0.753$\pm$0.228     &    -0.096$\pm$0.239     &     0.083$\pm$0.225     &     0.639$\pm$0.129     \\
       1.3-1.5       &    -0.354$\pm$0.002     &    -1.554$\pm$0.007     &    -0.524$\pm$0.009     &     0.057$\pm$0.010     &     0.128$\pm$0.092     &     0.586$\pm$0.104     \\

\hline
\multicolumn{7}{|c|}{ LMC} \\
\hline
        0.5-0.7       &    -0.356$\pm$0.014     &    -1.679$\pm$0.412     &    -0.843$\pm$0.450     &    -0.231$\pm$0.495     &    -4.434$\pm$0.412     &     0.000$\pm$0.000     \\
       0.7-0.9       &    -0.350$\pm$0.017     &    -1.329$\pm$0.353     &    -0.600$\pm$0.423     &    -0.169$\pm$0.465     &     0.175$\pm$0.963     &     0.658$\pm$1.120     \\
       0.9-1.1       &    -0.399$\pm$0.023     &    -1.346$\pm$0.398     &    -0.608$\pm$0.472     &    -0.138$\pm$0.494     &    -0.070$\pm$0.672     &     0.092$\pm$0.863     \\
       1.1-1.3       &    -0.495$\pm$0.016     &    -1.665$\pm$0.399     &    -0.711$\pm$0.478     &    -0.088$\pm$0.510     &     0.193$\pm$0.495     &     0.531$\pm$0.532     \\
       1.3-1.5       &    -0.498$\pm$0.010     &    -1.592$\pm$0.207     &    -0.518$\pm$0.257     &    -0.026$\pm$0.269     &     0.031$\pm$0.232     &     0.505$\pm$0.151     \\

\hline
\hline
$\log(P)$ &  $\Delta \phi_{31}(V,I)$  &   $\Delta \phi_{31}(V,K_{s})$ &     $\Delta \phi_{31}(J,K_{s})$ &   $\Delta \phi_{31}(H,K_{s})$ & $\Delta \phi_{31}(3.6,K_{s})$ &  $\Delta \phi_{31}(3.6,4.5)$ \\
\hline
\hline
\multicolumn{7}{|c|}{ Galaxy} \\
\hline
        0.5-0.7       &    -0.634$\pm$0.010     &    -2.251$\pm$0.386     &    -0.823$\pm$0.467     &     0.193$\pm$0.501     &     0.169$\pm$0.722     &     0.082$\pm$0.669     \\
       0.7-0.9       &    -0.621$\pm$0.013     &    -2.321$\pm$0.329     &    -0.812$\pm$0.392     &     0.057$\pm$0.417     &     0.272$\pm$0.540     &     0.293$\pm$0.574     \\
       0.9-1.1       &     0.418$\pm$0.012     &    -1.065$\pm$0.396     &    -0.753$\pm$0.486     &    -0.091$\pm$0.512     &     1.197$\pm$1.012     &     1.747$\pm$1.104     \\
       1.1-1.3       &    -0.591$\pm$0.007     &    -3.377$\pm$0.235     &    -1.673$\pm$0.284     &    -0.121$\pm$0.329     &     0.704$\pm$0.330     &     1.197$\pm$0.321     \\
       1.3-1.5       &    -0.700$\pm$0.002     &    -3.042$\pm$0.012     &    -1.104$\pm$0.014     &     0.116$\pm$0.017     &     0.198$\pm$0.178     &     1.229$\pm$0.211     \\

\hline
\multicolumn{7}{|c|}{ LMC} \\
\hline
       0.5-0.7       &    -0.633$\pm$0.027     &    -1.834$\pm$0.680     &    -0.557$\pm$0.787     &     0.018$\pm$0.826     &    -4.311$\pm$0.679     &     0.000$\pm$0.000     \\
       0.7-0.9       &    -0.678$\pm$0.032     &    -1.425$\pm$0.587     &    -0.151$\pm$0.760     &     0.363$\pm$0.776     &     0.390$\pm$1.410     &     3.220$\pm$1.748     \\
       0.9-1.1       &    -0.175$\pm$0.036     &    -1.023$\pm$0.609     &     0.047$\pm$0.782     &     0.521$\pm$0.785     &    -0.579$\pm$1.298     &     0.892$\pm$1.699     \\
       1.1-1.3       &    -0.791$\pm$0.022     &    -2.490$\pm$0.592     &    -0.687$\pm$0.720     &     0.457$\pm$0.781     &     0.951$\pm$0.852     &     1.561$\pm$0.772     \\
       1.3-1.5       &    -0.819$\pm$0.015     &    -1.691$\pm$0.381     &     0.304$\pm$0.447     &     0.640$\pm$0.504     &     0.392$\pm$0.444     &    -0.318$\pm$0.342     \\

\hline
\end{tabular}}
\end{center}
\end{table*}

\begin{figure}
\begin{center}
\includegraphics[width=0.5\textwidth,keepaspectratio]{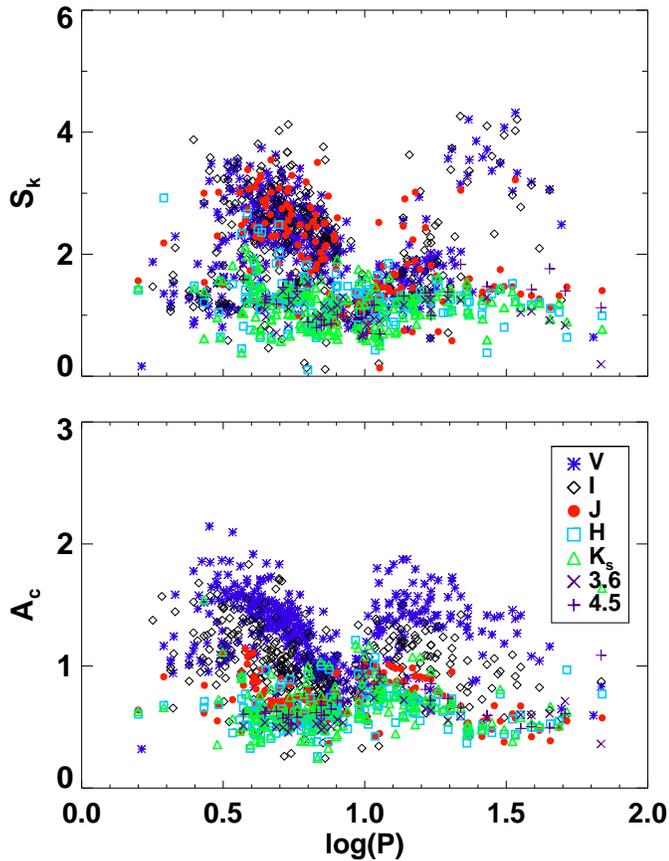}
\end{center}
\caption{ Variation of skewness ($S_{k}$) and acuteness ($A_{c}$) parameter against $\log(P)$ for Galactic Cepheids.} 
\label{fig:sk_ac_mw.eps}
\end{figure}

\begin{figure}
\begin{center}
\includegraphics[width=0.5\textwidth,keepaspectratio]{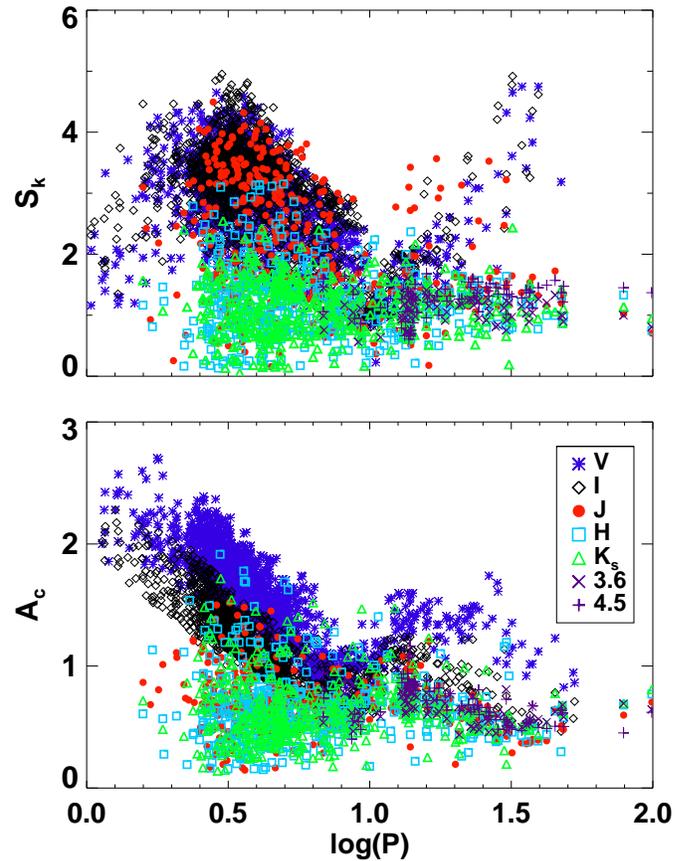}
\end{center}
\caption{ Same as Fig.~\ref{fig:sk_ac_mw.eps}, but for LMC Cepheids..} 
\label{fig:sk_ac_lmc.eps}
\end{figure}

The corresponding plots for LMC Cepheids are presented in
Fig.~\ref{fig:mean_fou_lmc.eps}. The same patterns present in Galactic
Cepheids are also seen in this sample. The data suggest a separation between
optical and infrared Fourier phase parameters for $\log(P) < 0.5$, which may
extend to $R_{21}$ but is not visible in $R_{31}$.
Fig.~\ref{fig:mean_fou_mw.eps} and \ref{fig:mean_fou_lmc.eps} were used to determine
the average behavior of the Fourier parameters with wavelength at given period. Flat sections in these plots do occur
when Fourier parameters oscillate from a high to low value or vice-versa, particularly when we are near the center of the HP. For 
example, the flat section in $R_{31}$ for LMC Cepheids in Fig.~\ref{fig:mean_fou_lmc.eps} is due to the $R_{31}$ 
clump at periods $0.8 <\log(P) < 1.0$ observed at optical bands in Fig.~\ref{fig:lmc_fou_lamda_mir.eps}. 
However, such plots do provide evidence that the Hertzsprung progression is most dramatic at shorter
wavelengths and in $R_{21}$ and $\phi_{21}$ parameters \citep{slee81}.

We quantitatively analyzed the progression of mean Fourier
parameters with period and wavelength by calculating the change in
mean parameter values (binned every 0.2~dex in $\log(P)$) across two
bands. We restricted the analysis to $0.5 < \log(P) < 1.5$ because this
period range provides smooth progressions for each parameter with
reduced scatter. The result of this analysis is given in Table~\ref{table:diff_fparams}, for both the Galaxy and LMC.
Comparing the $V$- and $I$-band results, we observed a negligible change in
amplitude parameters while there was a nearly constant offset
in phase parameters for all period bins except for the one centered at
$\log(P)=1.0$.
Comparing the optical to near-infrared results, the change in
amplitude parameters is small around $\log(P)=1.0$, increasing slowly up to
$\log(P)=1.3$ and sharply afterwards. The change in the values of amplitude 
parameters when comparing wavelengths
shorter and longer than $J$ is greatest for $1.3 < \log(P) < 1.5$.
In case of the phase parameters, we observed a similar and significant difference in most of the periods bins.
The comparisons of $H$ to $K_{\rm s}$ and 3.6 to 4.5-$\mu m$ exhibit
a small change in amplitude parameters and a large scatter in the
phase parameters.
We note that $\Delta\phi$ values 
for $(V,K_{\rm s}), (J,K_{\rm s}), (H,K{\rm_s})$  increase as a function of
wavelength, while $\Delta R$ values decrease as a function of
wavelength for $0.5 < \log(P) < 1.5$.

\subsection{Skewness and acuteness parameters}

We also observed the variation of skewness ($S_{k}$) and acuteness ($A_{c}$) parameters following the work of
\citet{stelling86, stelling87} and \citet{bonohp}. \citet{stelling87} defined skewness as the ratio of the phase duration of the 
descending branch to the phase duration of the rising branch. They defined acuteness as the ratio of the 
phase duration during which the magnitude is fainter than the median magnitude to the phase duration during which it is brighter
than median magnitude. If $\phi_{min}$ and $\phi_{max}$ are the phases corresponding to the extremum of the rising branch, 
the phase duration of the rising branch is $\phi_{rb} = \phi_{max} -\phi_{min}$. Similarly, following \citet{bonohp}, 
we defined the median magnitude to be, $m_{med} = 0.5\times(m_{max} + m_{min})$ and $\phi_{fw}$ as the full width at 
half maximum of the light curve, which is equivalent to phase duration of brighter than average light. Hence

\begin{equation*}
S_{k} = \frac{1}{\phi_{rb}} - 1;~~~~~~
A_{c} = \frac{1}{\phi_{fw}} - 1.
\end{equation*}

The skewness is a measure of left/right asymmetry and it 
decreases when the slope of the rising branch becomes flatter while acuteness is a
measure of the top-down asymmetry of the light curve and it decreases when the shape changes from 
sawtooth to flat-topped \citep{bonohp}. For observed stars, the $S_{k}$ is generally greater than unity while for
symmetric light curves both parameters attain a value close to 1.
Since, both skewness and acuteness parameters are a function of phase durations, we use equation~\ref{eq:foufit1}
to obtain 1000 data points per light curve to determine an accurate value of $\phi_{rb}$ and $\phi_{fw}$.  
The variation of $S_{k}$ and $A_{c}$ with period and wavelength is shown in Fig.~\ref{fig:sk_ac_mw.eps}
and \ref{fig:sk_ac_lmc.eps} for the Galaxy and LMC, respectively. At a fixed period, we find that the value of 
$S_{k}$ and $A_{c}$ decreases with wavelength. Also, the separation in the values of both parameters
for wavelengths shorter/longer than $J$-band is clearly visible for Cepheids having $\log(P) > 1.3$. This behavior is similar
to that of mean Fourier parameters discussed in the previous subsection. As the light curves from 
optical to infrared bands become more sinusoidal and flat-topped, both the parameters are generally
expected to decrease with wavelength. However, the Cepheids having period in the vicinity of 10 days are
more symmetric as both parameters attain a value close to unity.
We emphasize that the skewness/acuteness 
parameters are the functions of light curve shape similar to Fourier 
parameters so either set can be used to see the variation 
as they are not independent of each other.

\begin{figure}
\begin{center}
\includegraphics[width=0.5\textwidth,keepaspectratio]{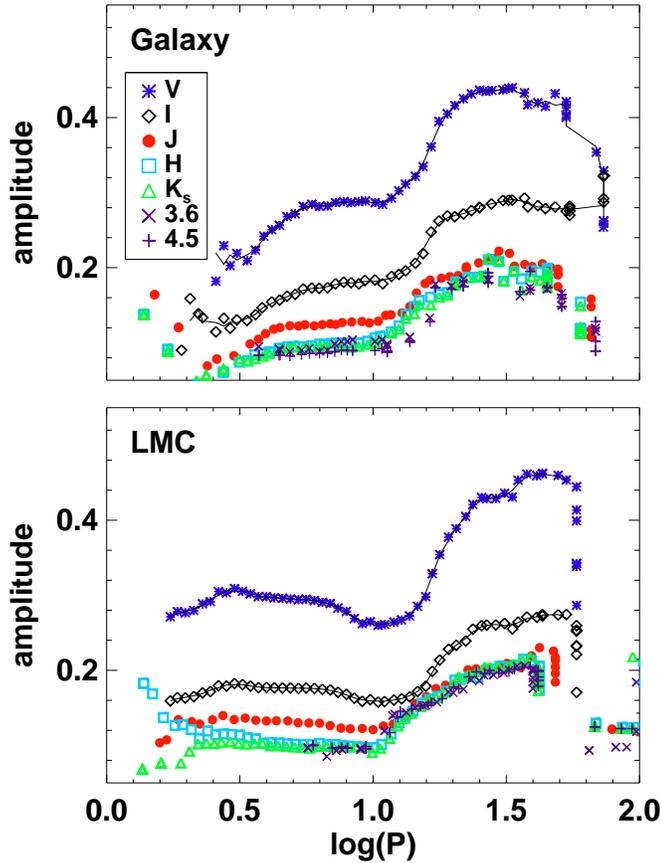}
\end{center}
\caption{Variation of the mean of light curve amplitudes in multiple bands with period for Galactic and LMC Cepheids.}
\label{fig:mean_amp_logp.eps}
\end{figure}

\section{The Variation of the Hertzsprung Progression with Wavelength} 
\label{sec:hp_logp}

We also observed the variation of mean of light curve amplitudes with period and wavelength.
We apply sliding mean calculations to determine the mean amplitudes similar to mean Fourier parameters.  
The variation of mean amplitudes at multiple bands for both Galactic and LMC Cepheids is shown 
in Fig.~\ref{fig:mean_amp_logp.eps}. The mean amplitudes decrease with increasing wavelengths. The
amplitudes in the optical bands show a sharp rise for periods $1.0 < \log(P) < 1.5$ as compared to infrared bands. 

\begin{figure}
\begin{center}
\includegraphics[width=0.5\textwidth,keepaspectratio]{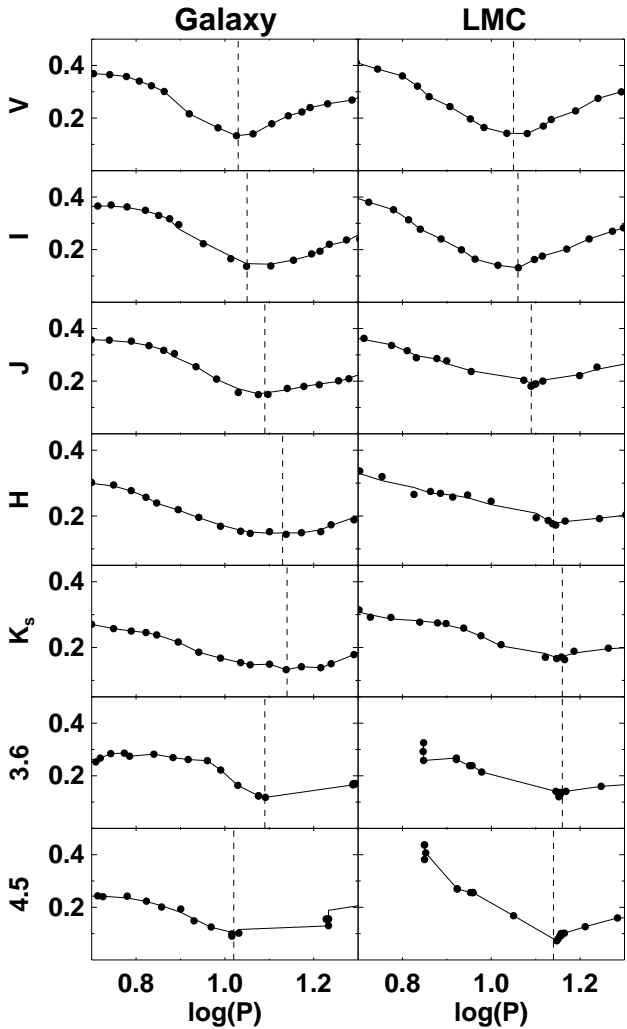}
\end{center}
\caption{Functional fits to mean Fourier amplitude parameter (R$_{21}$) in multiple-bands, used to determine the central period of HP.
The dashed vertical line represents the central period of HP in each band. Similarly, functional fits were also applied to mean light curve amplitudes to obtain the center of HP.}
\label{fig:r21_mean_fit.eps}
\end{figure}

\begin{figure}
\begin{center}
\includegraphics[width=0.5\textwidth,keepaspectratio]{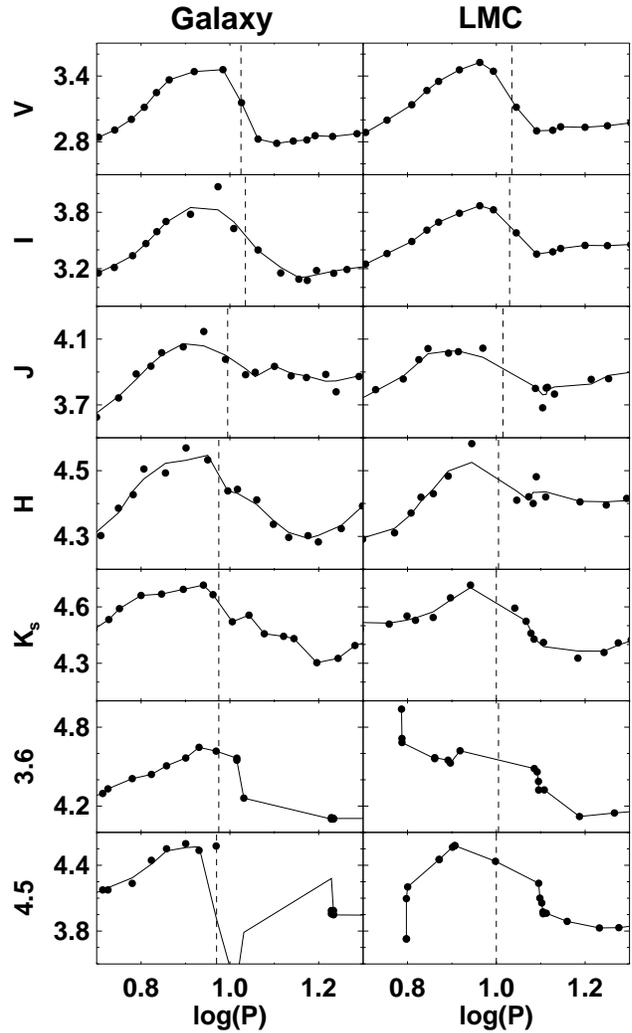}
\end{center}
\caption{Functional fits to mean Fourier phase parameter ($\phi_{21}$) in multiple-bands, used to determine the central period of HP.
The dashed vertical line represents the central period of HP in each band. Similarly, functional fits were also applied to  $\phi_{31}$ to 
obtain the center of HP.}
\label{fig:phi21_mean_fit.eps}
\end{figure}

The mean parameter plots for both Galactic and LMC Cepheids provide evidence for clearly visible trends that could be 
fit using functional forms.  We therefore reduced the step size in the 
sliding mean calculation to 0.02 in $\log(P)$ with the same bin width of $\log(P)=0.2$ and fit polynomials of 
varying degrees. These were then interpolated to obtain values every 0.01~dex. We have presented the functional fit
to $V$-band parameters in Figs.~\ref{fig:mean_fou_mw.eps}, \ref{fig:mean_fou_lmc.eps}  $\&$ \ref{fig:mean_amp_logp.eps}.
We also provide functional fits to multiple band Fourier amplitude parameter ($R_{21}$) and Fourier phase parameter
($\phi_{21}$). These plots will be used to determine the central period of HP, and are shown in 
Figs.~\ref{fig:r21_mean_fit.eps} \& \ref{fig:phi21_mean_fit.eps}. Similar functional fits were also applied to
light curve amplitudes and $\phi_{31}$ parameter to determine central period of HP.

\begin{figure}
\begin{center}
\includegraphics[width=0.5\textwidth,keepaspectratio]{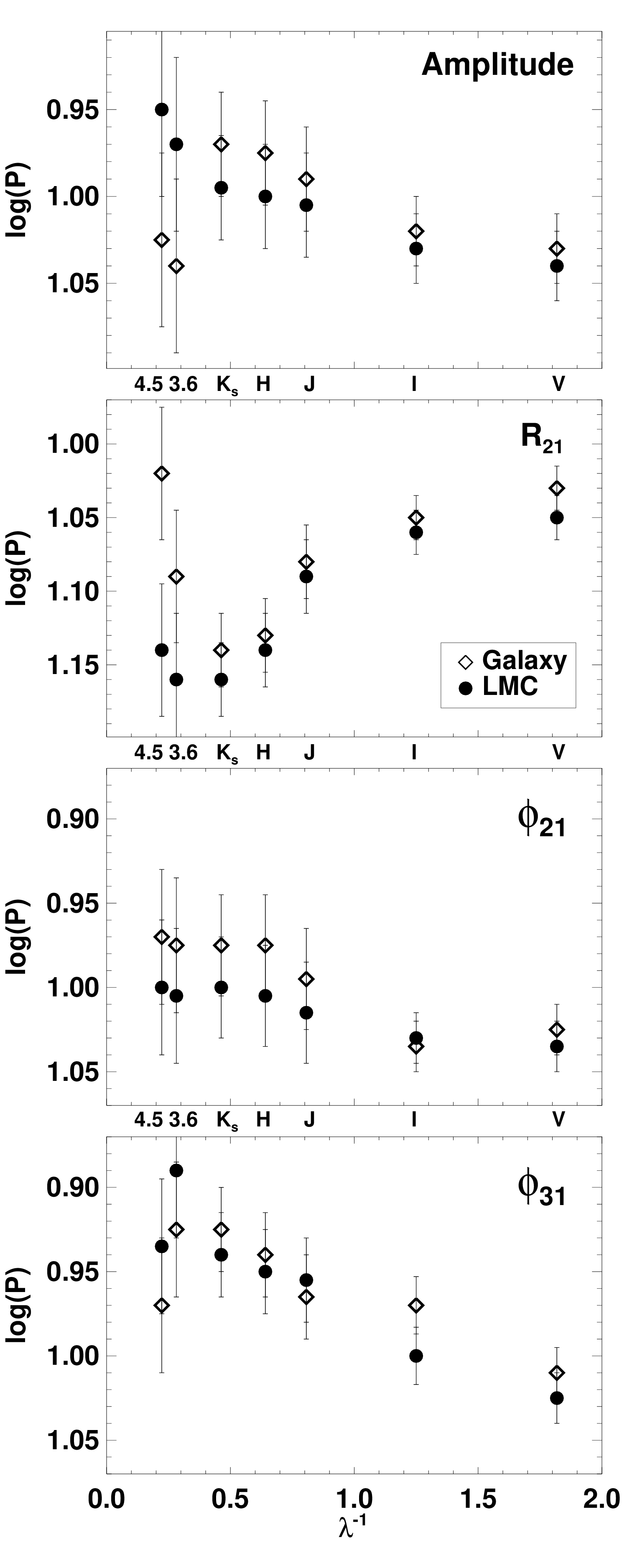}
\end{center}
\caption{Variation of central period of HP with wavelength for amplitude and Fourier parameters $R_{21}$, $\phi_{21}$ \&  $\phi_{31}$.
The error bars represent the maximum possible deviation from the median value with different degree of polynomials.}
\label{fig:hp_logp.eps}
\end{figure}

Following the work of \citet{bonohp}, we determine the central period of HP using light curve amplitudes.
However, we do not observe a sharp minima around $\log(P)=1.0$ but we also note that 
both theoretical light and velocity curve display a flatter minima in \citet{bonohp}.
The variation of the central period of the HP determined using the light curve amplitudes is
presented in the top panel of Fig.~\ref{fig:hp_logp.eps}. For optical wavelengths, 
we find that the central period of the HP is at $\log(P) = 1.04$ for $V$-band and
$\log(P) = 1.03$ for $I$-band, in the LMC. These results are in good agreement with the 
theoretical prediction of $\log(P)=1.051\pm0.018$ by \citet{bonohp}.

We determined the minimum values of $R_{21}$ as a function of wavelength for
each sample - these are given in the second panel of Fig.~\ref{fig:hp_logp.eps}.
The results argue for a clear trend in the central period of the HP, with the central
value increasing with wavelength. For Galactic data at wavelengths longer than $K_s$, the central period of
the HP shifts toward shorter periods.
The increased scatter at mid-infrared
wavelengths is expected as there are not enough stars in those bands, specially
in the vicinity of 10 days. We find that the central period of HP is at $\log(P) = 1.05$ for $V$-band and
$\log(P) = 1.06$ for $I$-band, in the LMC. Again, these results are in excellent agreement with those 
predicted using theoretical light curves \citep{bonohp} \& obtained using Fourier parameters \citep{welchhp97}.
However, we emphasize that the results of \citet{bonohp} are obtained using the amplitudes of 
light and velocity curves while our results are obtained using Fourier amplitude parameter.
So the central period of HP determined using the two methods are different but the agreement in
these two results is very interesting.
We do not plot the corresponding values for $R_{31}$
since the data imply that this parameter seems to be less sensitive to the bump progression and a shallow minimum is observed in 
Fig.~\ref{fig:mean_fou_lmc.eps}.

In the case of phase parameters, the break in the center of the
HP required a slightly different approach. 
We fit two polynomials to the points on each side of $\log (P)=1.0$, restricting
the range to $\pm0.2$~dex from that value. We then estimated the maximum value
before 10 days and the minimum value just after 10 days. We linearly interpolated
across these extrema and took the mid-point of the resulting line to be
the center of the HP. We do not observe sharp minima after 10 days at infrared bands
due to smaller amplitudes and larger scatter in phase parameters, as shown in 
Fig.~\ref{fig:phi21_mean_fit.eps}. However, there is a small but significant drop 
in the value of $\phi_{21}$ in the vicinity of 10 days.  In such cases, where the minima 
is flat or extended towards a longer period, we have chosen the first point lying on the 
functional fit as minima after 10 days. We consider the mid point of these extremums 
obtained from functional fits as the center of HP.
The results obtained from the fits to phase parameters are shown in the bottom two panels of 
Fig.~\ref{fig:hp_logp.eps}. Again, there seems to 
be a clear trend in the value of the center of the HP, 
but now decreasing with increasing wavelength. In case of 
$\phi_{21}$, the central period occurs at $\log(P) \sim 1.04$ in the
optical bands for both Galaxy and LMC, consistent with the previous
determination of $\log(P)=1.049\pm0.031$ by \citet{welchhp97}. 

We have also observed a slight difference in the central period of the
HP for the Galaxy and LMC in each parameter. 
This difference is most likely due to metallicity differences
between the two galaxies. 
As seen in Fig.~\ref{fig:hp_logp.eps}, the greatest disparity in the central period
of the HP happens longwards of $K_s$ for amplitude \& $R_{21}$ and beyond $J$ for
$\phi_{21}$. No significant difference is seen for $\phi_{31}$.
Since we have obtained these results using the sliding mean calculations, it is difficult to determine 
the exact significance of these results specially at longer wavelengths, where the number of stars is smaller
in the vicinity of $\log(P)=1.0$. Since we applied the same procedure 
at all bands, there is a differential effect in these parameters that seems to be real. 
Also, this confirms the work of \citet{beauli98}, who has determined the center of the HP for our
Galaxy, LMC and SMC Cepheids using Fourier parameters. \citet{beauli98} found a shift in the HP center towards 
the longer periods for the Galaxy having lower mean metallicity, following the work of \citet{ander87, ander88}. 
However, we emphasize that more data
will be needed to determine the central period more accurately, particularly at longer wavelengths.

\section{Conclusions}
\label{sec:discuss}

In the present study, we discussed the Fourier decomposition of Galactic and LMC
Cepheid light curves in multiple bands.  We compiled and made use of the largest
data sets available in each band.  We analyzed the variation of Fourier
parameters in detail to observe some interesting patterns. We found an increase
in phase parameters with increasing wavelength for both Galactic and LMC
variables. We also observed a decrease in amplitude parameters with
increasing wavelength. An interesting pattern in amplitude parameters
was observed, which suggests that for $VIJ$-band, the amplitude
parameters increases sharply as compared to longer wavelengths for periods greater than around 20 days.
Quantitatively this was summarized by determining the difference of mean Fourier parameters in multiple bands.
We also observed a decrease in skewness and acuteness parameters as function of wavelength at a fixed period
suggesting Cepheid light curves to be more symmetric at longer wavelengths.

The central period of the HP
displays a clear variation with increasing wavelength, suggesting an increase in central period for
Fourier amplitude parameters and a decrease for phase parameters. At optical bands, the mean central 
period of the HP occurs at $\log(P) \sim 1.03$ for the Galaxy and 
at $\log(P) \sim 1.04$ for the LMC, which are
consistent with the previous studies \citep{welchhp97, bonohp}. 
We also found small differences in the central period of the HP for
different Fourier parameters between the Galaxy and LMC. 
These differences are mainly in the $R_{21}$ beyond $K_s$ and in ${\phi}_{21}$  beyond $J$. 
These differences are such 
that Galactic data have the central period at shorter values. 
At optical bands, this difference is more accurate and confirms previous work by \citet{beauli98} but 
we can not determine the exact significance at infrared bands due to larger scatter in amplitude and phase
parameters.

\begin{figure}
\includegraphics[width=0.48\textwidth,keepaspectratio]{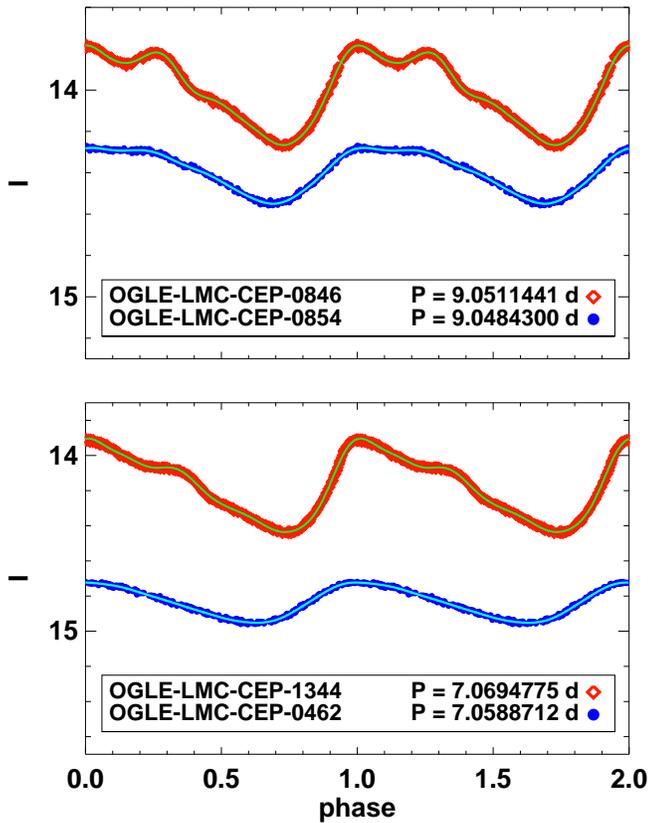}
\caption{Examples of I-band light curves of LMC Cepheids, which have periods in the vicinity of clump in $R_{31}$ parameters in 
Fig.~\ref{fig:lmc_fou_lamda_mir.eps}. Light curves in red/blue color are from in/out side of the clump respectively.} 
\label{fig:r31_bump_lc.eps}
\end{figure}

We also observed a flatter variation in $R_{31}$ amplitude parameter as compared to other Fourier parameters in the vicinity of 10 days.
This shallower minimum is more pronounced in the mean Fourier parameters shown in Fig.~\ref{fig:mean_fou_mw.eps} 
and Fig.~\ref{fig:mean_fou_lmc.eps}. Further, we observed a clump in the $R_{31}$ in the 
vicinity of $0.7 < \log(P) < 1.0$ for $VI$-bands (see Fig.~\ref{fig:lmc_fou_lamda_mir.eps}), 
which is a possible cause of flatter minimum in Fig.~\ref{fig:mean_fou_lmc.eps}. 
However, this clump is not visible for Galactic Cepheids because of the smaller number of
stars as compared to OGLE LMC data.   
We investigated the light curves of stars in/out side the clump but with similar periods. Examples are shown in 
Fig.~\ref{fig:r31_bump_lc.eps}. A slight bump after the maximum light (minimum magnitude) is observed in the light curves 
for the stars that are in the clump noted in the $R_{31}~vs.~\log(P)$ plot (Fig.~\ref{fig:lmc_fou_lamda_mir.eps}). 
This feature will be extensively studied in a future work
and may provide a direct link between light curve structure and Fourier parameters.

The variation of Fourier parameters with wavelength can shed light on pulsation
physics that are wavelength dependent. Further, these results can serve as a
benchmark to constrain theoretical stellar pulsation models that now
routinely incorporate model stellar atmospheres and produce light curves at various
wavelengths.

While a physical interpretation of Fourier parameters is still an open question, the method does
provide a quantitative description of the structure of Cepheid and RR Lyrae light curves.
In order to have more confidence in these models, it will be important
to compare these model and observed light curves, quantitatively with Fourier
parameters as a function of metallicity, wavelength and period.

\section*{Acknowledgments}
\label{sec:ackno}
AB is thankful to the Council of 
Scientific and Industrial Research, New Delhi, for a Junior Research 
Fellowship (JRF) and to Sukanta Deb for helpful discussions. This work is supported by the grant provided by 
Indo-U.S. Science and Technology Forum under the Joint Center for Analysis of Variable Star Data.
CCN thanks the funding from Ministry of Science and Technology (Taiwan) under
the contract NSC101-2112-M-008-017-MY3. LMM acknowledges support by the United States 
National Science Foundation through AST grant number 1211603 and by Texas A\&M 
University through a faculty start-up fund and the Mitchell-Heep-Munnerlyn 
Endowed Career Enhancement Professorship in Physics or Astronomy.
We also thank the anonymous referee for his/her useful suggestions that improved the quality
of the paper. This research was supported by the Munich Institute for Astro- and Particle Physics 
(MIAPP) of the DFG cluster of excellence ``Origin and Structure of the Universe''.
This work has made use of NASA's Astrophysics Data System, SIMBAD database and the VizieR catalogue, 
McMaster Cepheid Photometry database and the data products from the Two Micron All Sky Survey. 

\bibliographystyle{mn2e}
\bibliography{new_multi_fou}

\end{document}